\documentclass[amsmath,12pt,amssymb,preprint,nofootinbib]{revtex4-2}
\usepackage{amsfonts} 
\usepackage{graphicx} 
\usepackage{epsfig}
\usepackage{multirow}
\usepackage{bm}
\usepackage{color}
\usepackage{amssymb}
\usepackage{dsfont}
\usepackage{hyperref}
\usepackage{cleveref}
\newcommand{\be}{\begin{eqnarray}}
	\newcommand{\ee}{\end{eqnarray}}

\newcommand{\bfy}{{\bf y}_{\perp}}
\newcommand{\bfz}{{\bf 0}_{\perp}}

\newcommand{\bfk}{{\bf k}_{\perp}}

\newcommand{\bfkj}{{\bf k}_{\perp j}}

\newcommand{\bfP}{{\bf P}_{\perp}}

\begin{document}
\title{Effect of Asymmetric Nuclear Medium on the Valence Quark Structure of the Kaons}
\author{Dhananjay Singh}
\email{snaks16aug@gmail.com}
\author{Satyajit Puhan}
\email{puhansatyajit@gmail.com}
\author{Navpreet Kaur}
\email{knavpreet.hep@gmail.com}
\author{Manpreet Kaur}
\email{ranapreeti803@gmail.com}
\author{Arvind Kumar}
\email{kumara@nitj.ac.in}
\author{Suneel Dutt}
\email{dutts@nitj.ac.in}
\author{Harleen Dahiya}
\email{dahiyah@nitj.ac.in}
\affiliation{Department of Physics, Dr. B.R. Ambedkar National
	Institute of Technology, Jalandhar, 144008, India}

	\begin{abstract}
		The role of asymmetric nuclear medium on the properties of kaon is investigated at zero and finite temperature employing a hybrid approach integrating the light cone quark model (LCQM) and the chiral SU(3) quark mean field (CQMF) model. The in-medium quark masses are calculated within the CQMF model and are used as inputs to study the medium modifications in the kaon properties. In particular, we have analysed the impact of baryonic density, isospin asymmetry and temperature on the weak decay constant, distribution amplitudes (DAs) and parton quark distributions (PDFs) of valence quarks of kaon. The effects of isospin asymmetry on the kaon doublet $K =\left(\begin{array}{c} K^{+} \\ K^{0} \end{array} \right)$ and antikaon doublet $\bar{K}$= ($K^-, \bar{K}^0$) are also studied. In order to compare with future experiments, we have also evolved the in-medium DAs and PDFs of kaons to $Q^2=16$ GeV$^2$. As compared to the temperature and isospin asymmetry, change in baryonic density of the nuclear medium makes more significant changes to the DAs and PDFs of kaons.
		
	\end{abstract}
	
	\maketitle
\newpage	
	\section{\label{intro}Introduction}
Heavy nuclei collisions create a nuclear medium that affects the physical observability of the specific processes. This was first verified by the European Muon Collaboration (EMC) in the deep-inelastic scattering (DIS) experiment \cite{EuropeanMuon:1983wih}. They observed that the ratio of the DIS cross-sections for nuclei to those for deuterium deviate from unity, indicating that the parton distributions in bound nucleons are different from those in free nucleons. The EMC effect suggests influence of the surrounding nuclear medium on internal structure of nucleons which clearly  implies complex interaction of nucleons and their environment within the nucleus. This has crucial consequences for understanding strong force and the behavior of nuclear matter \cite{Geesaman:1995yd}. A small change resulting from medium modifications significantly affects the internal structure of hadrons as well as the meson photo productions in heavy ion collisions \cite{NA49:2007stj}. The reduction of pion decay constant in the nuclear medium has also been observed in deeply bound pionic atom experiments and pion-nucleus scattering \cite{Suzuki:2002ae}. Investigating the changes in hadron properties within dense nuclear environment is vital for comprehending the observables in high-energy physics experiments. This research is also essential for exploring the properties of quantum chromodynamics (QCD), such as the breaking of chiral symmetry at low energies and its restoration under conditions of high density and temperature.

\par Pions and kaons are the lightest pseudo-scalar mesons used to examine the influence of the nuclear dense medium on their constituent quarks at low energy QCD \cite{Ioffe:2005ym}. At low energy $Q^2 \le 1$ GeV$^2$, the valence quark contributions dominate over the gluon and sea quark contributions. Therefore, it becomes essential to study the valence quark properties like parton distribution functions (PDFs) in the nuclear medium at this scale. However, it can be evolved to higher $Q^2$ through an evolution tool kit for the comparisons with experimental results. PDFs are the one-dimensional distribution functions that carry the information about the longitudinal momentum fraction ($x$) carried by a quark from its parent hadron and the experimental data of these PDFs can be extracted from DIS experiments \cite{Alekhin:2002fv,Polchinski:2002jw}. Experimental data \cite{Bourrely:2020izp,Aicher:2010cb}, lattice simulations \cite{Alexandrou:2021mmi}, and model calculations excluding any medium \cite{Puhan:2023ekt,Pasquini:2014ppa,Cao:1997sx} and in-medium pion PDFs \cite{Hutauruk:2019ipp,Puhan:2024xdq} have been investigated extensively. Meanwhile, distribution amplitudes (DAs) are studied to understand the quark-antiquark interactions and coupling strength inside a hadron. It quantifies the likeliness amplitude for locating these constituent particles to bound together, thereby interpreting the effectiveness of their
interactions in forming the structure of a hadron \cite{Brodsky:1989pv}. 

\par 
Studies of kaon photoproduction on nuclei \cite{Lee:1999kd} and kaon-nucleus Drell-Yan processes \cite{Londergan:1996vh} were aimed to uncover the internal structure of kaons in the nuclear medium. In recent times, the quark meson coupling (QMC) model has been combined with the light-front constituent quark model (LCQM) to study the medium-modified electromagnetic form factors of kaons \cite{Yabusaki:2017sgs}. Additionally, the QMC model has also been used with the Nambu-Jona-Lasino (NJL) model to investigate the valence PDFs of pions and kaons in symmetric nuclear medium \cite{Hutauruk:2018qku}. In Ref. \cite{Arifi:2023jfe}, the light-front quark model (LFQM) and QMC model were combined to explore the DAs and weak decay constants of the pseudoscalar as well as vector mesons in symmetric nuclear matter. The medium-modified elastic electromagnetic form factors of pion and kaon have also been explored using the Schwinger proper-time NJL model \cite{Gifari:2024ssz}. All these studies employ a hybrid approach of utilizing the QMC model to compute the in-medium properties of quarks and input them to investigate the in-medium properties of mesons. Notably, in these works, the medium modifications of meson properties have been explored in the symmetric nuclear medium at zero temperature. Recently, the spectroscopic properties of kaon at finite temperature and baryon chemical potential were explored using the QCD sum rules \cite{Bozkir:2022lyk}. 

\par In the present study, we aim to investigate the medium-modified DAs and PDFs of kaons in isospin asymmetric nuclear matter at finite temperatures. We compute the medium-modified DAs and PDFs of kaons in the LCQM by using the in-medium masses of quarks calculated in the CQMF model at finite temperature ($T$) and isospin asymmetry ($\eta$). We have also derived Mellin moments from the PDFs, which are connected to structure functions and are accessible in DIS processes. In an isospin symmetric medium, the $u$ and $d$ quarks are treated on the same footing, i.e., their properties, such as their in-medium masses, are equal. However, in heavy-ion collision experiments and inside the neutron stars, the bulk matter is found to be isospin asymmetric \cite{Chu:2012rd}. Hence, the isospin-dependent medium effects are crucial in real-world scenarios. In light of this, it becomes essential to study the effects of isospin asymmetry on the kaon doublet $K =\left(\begin{array}{c} K^{+} \\ K^{0} \end{array} \right)$ and antikaon doublet $\bar{K}$ = ($K^-, \bar{K}^0$). Many theoretical \cite{Mishra:2008dj,Li:1997zb,Lutz:1997wt} and experimental \cite{KaoS:1999gal,KaoS:2000eil,KAOS:2000ekm} studies have been carried to explore the isospin doublets of $K$ and $\bar{K}$.

\par We introduce the effect of isospin asymmetry on the in-medium properties of kaons through the chiral SU(3) quark mean field (CQMF) model \cite{Wang:2001jw}. This model treats quarks as degrees of freedom and binds them within hadrons through a confining potential. In this model, the interaction among quarks is described through the scalar meson fields $\sigma$, $\zeta$, and $\delta$, which are responsible for the attractive part of the interactions and the vector meson fields $\omega$ and $\rho$ accounting for the repulsive interactions in the isospin asymmetric nuclear medium. The model employs a mean-field approximation, meaning that the meson fields are treated as classical fields. As a result, only scalar and vector fields contribute to the quark-meson interaction Lagrangian. 
The CQMF model has been used to examine the properties of nuclear matter \cite{Wang:2001jw} as well as strange hadronic matter \cite{Wang:2001hw}. The magnetic moments of octet and decuplet baryons at finite temperature have also been studied with the CQMF model, both in nuclear \cite{Singh:2016hiw,Singh:2017mxj,Tsushima:2020gun} as well as strange hadronic matter \cite{Singh:2020nwp,Singh:2018kwq,Kumar:2023owb}. The CQMF model can also be utilized to study the mixed phase, wherein both hadrons and quarks coexist \cite{Wang:2006fh}. Recently, we have applied the CQMF model in conjunction with the LCQM to investigate the in-medium DAs and PDFs \cite{Puhan:2024xdq} as well as transverse momentum-dependent parton distributions (TMDs) of pions \cite{Kaur:2024wze}. 
LCQM works on the baseline of light-cone dynamics and provides a non-perturbative approach to study the hadron structure relativistically \cite{Zhang:1994ti}. This framework is gauge invariant in nature and primarily focuses on the valence quarks of the hadrons, which are the key constituents for defining the overall structure and properties of hadrons.  The light-cone wavefunctions have the ability to evaluate any hadronic quantity by convolution with the appropriate quark and gluon matrix elements \cite{Brodsky:1997de}. Within this model, comparable results have been found for available experimental data of PDFs \cite{Xiao:2002iv} and other theoretical predictions for multi-dimensional distribution functions \cite{Kaur:2019jow}. 

\par The outline of the paper is as follows: In Sec. \ref{cqmf}, we briefly review the CQMF model used in the present investigation to compute the in-medium masses of quarks. Sec. \ref{lcqm} presents the implementation of the in-medium masses into the LCQM model to obtain medium-modified properties of kaons in asymmetric nuclear medium. The medium-modified properties of kaon, such as their DAs and PDFs, have been explored in Sec. \ref{distribution} and Sec. \ref{parton}. In Sec. \ref{results}, we discuss the results obtained for the DAs and PDFs. Sec. \ref{summ} summarizes and concludes the present investigation.

\section{Models}
\subsection{\label{cqmf}Chiral SU(3) quark mean field model}
The CQMF model, which is used to calculate the effective masses of quarks in the nuclear medium,  builds upon
low energy properties of QCD \cite{Wang:2001jw} and broken scale invariance  \cite{Kharzeev:2008br}
to describe interactions among quarks and mesons, as well as among mesons, across various temperature and density conditions. The quark-meson interactions are governed through the exchange of scalar ($\sigma$, $\zeta$ and $\delta$) and vector ($\omega$ and $\rho$) fields leading to modification in the properties of baryons expressed in terms of their constituent quark flavors within the asymmetric nuclear medium. The non-strange scalar-isoscalar field $\sigma$ is associated with the $f_0$ ($\sim$ 500 MeV) scalar meson which consists of light $(u,d)$ quarks and $(\bar{u},\bar{d})$ antiquarks as its content. The $\zeta$ field is a strange scalar-isoscalar field associated with the scalar meson containing strange quark content. Additionally, the isospin asymmetry of the medium is incorporated in the mean-field relativistic models through the inclusion of the scalar isovector field $\delta$ and vector isosvector field $\rho$, while the dilaton field $\chi$ is considered in this model to incorporate the broken scale invariance property of QCD \cite{Wang:2001jw}. 
\par
The thermodynamic potential for isospin asymmetric nuclear matter at finite temperature and density can be described as 
\begin{eqnarray}
	\Omega &=& -\frac{k_{B}T}
	{(2\pi)^3} \sum_{i} \gamma_i
	\int_0^\infty d^3k\biggl\{{\rm ln}
	\left( 1+e^{- [ E^{\ast}_i(k) - \nu_i^* ]/k_{B}T}\right) \nonumber \\
	&+& {\rm ln}\left( 1+e^{- [ E^{\ast}_i(k)+\nu_i^* ]/k_{B}T}
	\right) \biggr\} -{\cal L}_{M}-{\cal V}_{\text{vac}}.
	\label{Eq_therm_pot1}  
\end{eqnarray}
Here, the summation over  nucleons in the medium represented by $i$ = $p/n$ and the degeneracy factor $\gamma_i=2$ reflects two spin states of each nucleon. The effective energy of baryons $E_i^{\ast }(k)$ can be expressed as $E^{\ast }_i(k)=\sqrt{M_i^{\ast 2}+k^{2}}$, where $M_i^{\ast }$ is the effective mass of baryon. The term ${\cal V}_{\text{vac}}$ is subtracted to acquire zero vacuum energy. The effective chemical potential of baryon $\nu_i^\ast$ can be described in terms of free chemical potential $\nu_i$ as
\begin{eqnarray}
\nu_i^\ast = \nu_i - g_{\omega}^i\omega -g_{\rho}^i I^{3i} \rho,
\end{eqnarray}
where \(g_{\omega}^i\), and \(g_{\rho }^i\) are the coupling constants of the vector fields \(\omega\) and \(\rho\), respectively.

\par
The term, $\mathcal{L}_{M}$, in Eq.~(\ref{Eq_therm_pot1}) is mesonic Lagrangian, which can be written as \cite{Wang:2001hw, Wang:2002aq}  
 \begin{equation}
\label{mesonic_L}
\mathcal{L}_{M} 
  = {\cal L}_{\Sigma\Sigma}+ \mathcal{L}_{\text{VV}} + \mathcal{L}_{SB}.  \end{equation}
Here, the term ${\cal L}_{\Sigma\Sigma}$ represents the self-interaction of scalar mesons and in the mean-field approximation is expressed as \cite{Heide:1993yz}  
 \begin{eqnarray}
	{\cal L}_{\Sigma\Sigma} &=& -\frac{1}{2} \, k_0\chi^2
	\left(\sigma^2+\zeta^2+\delta^2\right)+k_1 \left(\sigma^2+\zeta^2+\delta^2\right)^2
	\nonumber \\ 
	&+&k_2\left(\frac{\sigma^4}{2} +\frac{\delta^4}{2}+3\sigma^2\delta^2+\zeta^4\right)+ k_3\chi\left(\sigma^2-\delta^2\right)\zeta \nonumber \\ 
	&-& k_4\chi^4-\frac14\chi^4 {\rm ln}\frac{\chi^4}{\chi_0^4} +
	\frac{\xi}
	3\chi^4 {\rm ln}\left(\left(\frac{\left(\sigma^2-\delta^2\right)\zeta}{\sigma_0^2\zeta_0}\right)\left(\frac{\chi^3}{\chi_0^3}\right)\right). \label{scalar0}
\end{eqnarray}
The scale-breaking effect is introduced by the last two logarithmic terms to compute the trace of the energy-momentum tensor within this model. 
The vector meson self-interactions term, ${\cal L}_{VV}$ is represented as
\begin{eqnarray}
	{\cal L}_{VV} =&\frac{1}{2} \, \frac{\chi^2}{\chi_0^2} \left(m_\omega^2\omega^2+m_\rho^2\rho^2\right)  + g_4\left(\omega^4+6\omega^2\rho^2+\rho^4\right) \, . 
	\label{vector}
\end{eqnarray}
\par
The Lagrangian density term, \(\mathcal{L}_{SB}\), is responsible for the explicit breaking of chiral symmetry, which generates non-zero masses for pseudoscalar mesons and can be expressed as \cite{Wang:2002aq, Papazoglou:1998vr}  
\begin{equation}\label{L_SB}
	{\cal L}_{SB}=\frac{\chi^2}{\chi_0^2}\left[m_\pi^2\kappa_\pi\sigma +
	\left(
	\sqrt{2} \, m_K^2\kappa_K-\frac{m_\pi^2}{\sqrt{2}} \kappa_\pi\right)\zeta\right] \, .
\end{equation}

The CQMF model employs a confinement mechanism for quarks within baryons through a Lagrangian term $\mathcal{L}_c = - \bar{\Psi} \chi_c \Psi$. Under the additional influence of meson mean fields, the Dirac equation governing a quark field $\Psi_{qi}$  is expressed as
\begin{equation}
	\left[-i\vec{\alpha}\cdot\vec{\nabla}+\chi_c(r)+\beta m_q^*\right]
	\Psi_{qi}=e_q^*\Psi_{qi}. 
 \label{Dirac}
\end{equation}
 Here, the subscripts $q$ and $i$ represent the quark $q$ (where $q = u, d, s$) within a baryon of type $i$ (where $i = p, n$). The effective quark mass $m_{q}^\ast$ and energy $e_{q}^\ast$ can be described by the following relations in terms of scalar ($\sigma, \zeta$ and $\delta$) and vector fields ($\omega$ and $\rho$) as
 \begin{equation}
	m_q^*=-g_\sigma^q\sigma - g_\zeta^q\zeta - g_\delta^q I_{3}^{q} \delta + \Delta m_q, \label{qmass}
\end{equation}
and 
\begin{equation} 
	e_q^*=e_q-g_\omega^q\omega-g_\rho^q I^{q}_3\rho\,,
	\label{eq_eff_energy1}
\end{equation}
respectively. The coupling constants of quarks with the scalar fields $\sigma$, $\zeta$, and $\delta$ are represented by $g_{\sigma}^q$, $g_{\zeta}^q$, and $g_{\delta}^q$, respectively. Whereas $I_3^q$ denotes the third component of isospin for the quark flavor and is given by $I_3^u=-I_3^d = 1/2$ and $I_3^s = 0$. The additional mass term $\Delta m_{u/d} = 0$ and $\Delta m_s = 77$ MeV are determined to fit the vacuum masses of quarks.
 The effective mass of the baryon $\mathcal{M}_i^{\ast }$,  is related to spurious center of momentum $\langle p_{i~cm}^{\ast 2} \rangle$ and effective quark energy $e_q^*$ as  \cite{Wang:2001jw}
\begin{eqnarray}
	\mathcal{M}_i^\ast = 
	\sqrt{\biggl(\sum_q n_{qi} e_q^\ast + E_{i~spin} \biggr)^2  - \langle p_{i~cm}^{\ast 2} \rangle} \,.
\end{eqnarray}
Here, $n_{qi}$ represents the number of $q$-flavored quarks in the $i^{th}$ baryon. The term $E_{i~spin}$ serves as a correction factor to baryon energy resulting from spin-spin interaction and is calibrated to reproduce the baryon vacuum mass.
The spurious center of momentum of baryon $\langle p_{i~cm}^{\ast 2} \rangle$ can be expressed in terms of $e_q^*$ and $m_q^*$ via following relation \cite{Barik:1985rm, Barik:2013lna} 

\begin{eqnarray}
    \langle p_{i,\text{cm}}^{\ast 2} \rangle = \frac{(11 e_q^* + m_q^*)}{6 (3 e_q^* + m_q^*)} \left( e_q^{\ast 2} - m_q^{\ast 2} \right).  
\end{eqnarray}

The total thermodynamic potential defined in Eq. (\ref{Eq_therm_pot1}) is minimized with respect to mesonic fields $\phi$, with $\frac{\partial \Omega}{\partial \phi} = 0$, to compute the coupled equations of motion of these fields. These equations are as follows  

\begin{align}
\label{Eq_sigma_eq1}
&k_0\chi^2\sigma-4 k_1\left(\sigma^2+\zeta^2+\delta^2\right) \sigma-2 k_2\left(\sigma^3+3 \sigma \delta^2\right) 
 -2 k_3 \chi \sigma \zeta-\frac{\xi}{3} \chi^4\left(\frac{2 \sigma}{\sigma^2-\delta^2}\right)  \nonumber\\ &+\left(\frac{\chi}{\chi_0}\right)^2 m_\pi^2 f_\pi + \left(\frac{\chi}{\chi_0}\right)^2 m_\omega \omega^2 \frac{\partial m_\omega}{\partial \sigma}-\left(\frac{\chi}{\chi_0}\right)^2 m_\rho \rho^2 \frac{\partial m_\rho}{\partial \sigma}
 =\sum_{i= p,n} g_{\sigma i} \rho_i^s ,
\end{align}
\begin{align}
& k_{0}\chi^{2}\zeta-4k_{1}\left( \sigma^{2}+\zeta^{2}+\delta^{2}\right)
\zeta-4k_{2}\zeta^{3}-k_{3}\chi\left( \sigma^{2}-\delta^{2}\right) 
-\frac{\xi}{3}\frac{\chi^{4}}{\zeta} \nonumber\\
&+\left(\frac{\chi}{\chi_{0}} \right)
^{2}\left[ \sqrt{2}m_{K}^{2}f_{K}-\frac{1}{\sqrt{2}} m_{\pi}^{2}f_{\pi}\right]
 =\sum_{i=p,n} g_{\zeta i}\rho_{i}^{s} ,
 \label{Eq_zeta11}
\end{align}
\begin{align}
\label{Eq_delta}
	&k_0 \chi^2 \delta-4 k_1\left(\sigma^2+\zeta^2+\delta^2\right) \delta-2 k_2\left(\delta^3+3 \delta \sigma^2\right)
 -2 k_3 \chi \delta \zeta \nonumber\\
 &-\frac{\xi}{3} \chi^4\left(\frac{2 \delta}{\sigma^2-\delta^2}\right)=\sum_{i=p,n} g_{\delta i}I_{3}^i \rho_i^s ,
\end{align}
\begin{align}
\label{Eq_chi}
&k_0 \chi\left(\sigma^2+\right.\left.\zeta^2+\delta^2\right)-k_3 \chi\left(\sigma^2-\delta^2\right) \zeta+\left[4 k_4+1-\ln \frac{\chi^4}{\chi_0^4}-\frac{4 d}{3} \ln \left(\frac{\left(\sigma^2-\delta^2\right) \zeta}{\sigma_0^2 \zeta_0}\right)\right]\chi^3\nonumber\\ &+ \frac{2 \chi}{\chi_0^2}\left[m_\pi^2 f_\pi \sigma+\left(\sqrt{2} m_K^2 f_K-\frac{1}{\sqrt{2}} m_\pi^2 f_\pi\right) \zeta\right]-\frac{\chi}{\chi_0^2}\left(m_\omega^2 \omega^2+m_\rho^2 \rho^2\right)=0 ,
\end{align}
\begin{align}
\label{Eq_omega_field}
	\frac{\chi^2}{\chi_0^2}\left(m_\omega^2 \omega\right)+4 g_4 \omega^3+12 g_4 \omega \rho^2=\sum_{i=p,n} g_{\omega i} \rho_i^v ,
\end{align}
\begin{align}
\label{Eq_rho_field}
	\frac{\chi^2}{\chi_0^2}\left(m_\rho^2 \rho\right)+4 g_4 \rho^3+12 g_4 \omega^2 \rho=\sum_{i=p,n} g_{\rho i}I_{3}^i \rho_i^v .
\end{align}
Here, $I_3^p=-I_3^n=1/2$ represents the third component of isospin for the nucleon. $m_\pi$, $m_K$, $m_\omega$ and $m_\rho$ represent the masses of the $\pi$, $K$, $\omega$ and $\rho$ mesons, respectively.
Additionally, $\rho_i^v$ and $\rho_i^s$ denote the vector and scalar densities of the $i^{th}$ baryon, respectively, and can be expressed as
\begin{eqnarray}
\rho_{i}^{v} = \gamma_{i}\int\frac{d^{3}k}{(2\pi)^{3}}  
\Bigg(\frac{1}{1+\exp\left[\beta(E^{\ast}_i(k) 
-\nu^{*}_{i}) \right]}-\frac{1}{1+\exp\left[\beta(E^{\ast}_i(k)
+\nu^{*}_{i}) \right]}\Bigg) ,
\label{rhov0}
\end{eqnarray}
and
\begin{eqnarray}
\rho_{i}^{s} = \gamma_{i}\int\frac{d^{3}k}{(2\pi)^{3}} 
\frac{m_{i}^{*}}{E^{\ast}_i(k)} \Bigg(\frac{1}{1+\exp\left[\beta(E^{\ast}_i(k) 
-\nu^{*}_{i}) \right]}+\frac{1}{1+\exp\left[\beta(E^{\ast}_i(k)
+\nu^{*}_{i}) \right]}\Bigg) .
\label{rhos0}
\end{eqnarray}
 The set of non-linear Eqs. (\ref{Eq_sigma_eq1})–(\ref{Eq_rho_field}) are solved for different values of baryon density, temperature, and isospin asymmetry of the medium. The parameters $\eta$ is defined as $-\frac{\sum_i I_{3}^{i} \rho^{v}_{i}}{\rho_{B}}$. Here, $\rho_B= \rho_p+\rho_n$ represents the total baryonic density of the nuclear medium.
\subsection{\label{lcqm} Light-cone quark model}
In the standard light-cone frame, we have chosen the four-vector notation as $y=[y^+,y^-,\bfy]$. If a meson with longitudinal spin projection $S_z$ has total momentum $P$, in terms of light-cone coordinates, $P$ can be expressed as \cite{Qian:2008px} 
\be 
P &=& \bigg(P^+,\frac{M^{\ast2}}{P^+},\bfz\bigg) \, , 
\ee
where $M^*$ represents the effective mass of a hadron.
The eigenstate of a meson $|\mathcal{H}(P^+,\bfP,S_z)\rangle$ can be expanded as a linear combination of multiparticle Fock eigenstates $|n\rangle$ as \cite{Lepage:1980fj} 
\be
|\mathcal{H}(P^+,\bfP,S_z)\rangle &=& \sum_{n,\lambda_j} \int \prod_{j=1}^{n} \frac{dx_j~  d^2\bfkj}{2(2\pi)^3\sqrt{x_{j}}} \, 16 \pi^{3} \, \delta \bigg(1-\sum_{j=1}^{n} x_{j}\bigg) \, \delta^{(2)} \bigg(\sum_{j=1}^{n}\bfkj\bigg) \nonumber \\		
&\times& \psi_{n/\mathcal{M}}(x_{j},\bfkj,\lambda_{j})|n; x_{j} P^{+},x_{j}\bfP + \bfkj,\lambda_{j}\rangle \, ,
\label{HadronState}\ee
where $\bfkj$ and $\lambda_j$ respectively denote the transverse momentum and helicity carried by the $j$th constituent parton of a hadron. The longitudinal momentum fraction associated with it is represented by $x_j$ that can be defined as $\frac{\textbf{k}_j^+}{P^+}$. The multiparticle state having $n$ number of particles is normalized as 
\be
\langle n; k^{\prime +}_j, \bfkj^\prime, \lambda_{j}^\prime|n ; k^+_j, \bfkj, \lambda_j \rangle = \prod_{j=1}^{n} 16 \pi^{3} \,  k^{\prime +}_j \, \delta (k^{\prime +}_j-k^+_j) \, \delta^{(2)} ( \bfkj^\prime-\bfkj) \, \delta_{\lambda_{j}^\prime \lambda_{j}} \, .
\ee
In the light-cone dynamics, according to restraint $\sum_{j=1}^{n} x_j = 1$ for the light-cone momentum fraction carried by a constituent quark and antiquark flavours of a meson, we have $x_1 + x_2 =1$ which states that if a quark carries $x$ fraction of longitudinal momentum, then the antiquark will carry the fraction $(1-x)$ from its parent meson. For the case of kaon, the total momenta carried by its constituent $u$ quark flavour and $\bar{s}$ antiquark flavor is represented by 
\be 
k_1 &=& \bigg(x P^+,\frac{\bfk^2 + m^{\ast2}_q}{x P^+},\bfk \bigg) \, ,  \nonumber \\
k_2 &=& \bigg((1-x) P^+,\frac{\bfk^2 + m^{\ast2}_{\bar{q}}}{(1-x) P^+},-\bfk \bigg) \, , 
\ee
where $m^*_q$ and $m^*_{\bar{q}}$ correspond to the effective masses of kaon's constituent quark and antiquark flavours, respectively. The two-particle Fock state of kaon having longitudinal spin projection $S_z=0$ can be written in terms of light-cone wave functions (LCWFs) as
\be 
|K (P^+,\bfP,S_z=0)\rangle &=& \int \frac{dx \, d^2 \bfk}{  16 \pi^3 \sqrt{x(1-x)}} \, \big[\psi_K (x,\bfk,\uparrow,\uparrow) \, |x P^+, \bfk, \uparrow, \uparrow \rangle   \nonumber \\
&+& \psi_K (x,\bfk,\uparrow,\downarrow) \, |x P^+, \bfk, \uparrow, \downarrow \rangle + \psi_K (x,\bfk,\downarrow,\uparrow) \, |x P^+, \bfk,  \downarrow,\uparrow \rangle \nonumber \\ &+& \psi_K (x,\bfk,\downarrow,\downarrow) \, |x P^+, \bfk, \downarrow, \downarrow \rangle \big] \,. \label{eqnq}
\ee 
These LCWFs can be obtained by taking the product of spin $\Phi_K$ and momentum space $\varphi_K$ wave functions as \cite{Huang:1994dy} 
\be 
\psi_K(x,\bfk,\lambda_1, \lambda_2)=  \Phi_K(x,\bfk,\lambda_1, \lambda_2) \, \varphi_K(x,\bfk) \, ,
\ee 
where $\lambda_1$ and $\lambda_2$ denote the helicity of the constituent quark and anti-quark of kaon. Light-cone spin state wave functions are obtained from instant-form by using Melosh Wigner rotation, and they can be expressed as \cite{Xiao:2002iv}
\be
\Phi_K(x,\bfk,\uparrow,\uparrow) &=& \frac{1}{\sqrt{2}\varUpsilon_1^{*} \varUpsilon_2^{*}} [(M^\ast x +m^\ast_q) k_2^l - (M^{\ast} (1-x) +m^\ast_{\bar{q}}) k_1^l] \, , \nonumber \\
\Phi_K(x,\bfk,\uparrow,\downarrow) &=& \frac{1}{\sqrt{2}\varUpsilon_1^{*} \varUpsilon_2^{*}} [(M^\ast x +m^\ast_q) (M^\ast (1-x)+m^\ast_{\bar{q}})-\bfk^2] \, , \nonumber \\
\Phi_K(x,\bfk,\downarrow,\uparrow) &=& \frac{-1}{\sqrt{2}\varUpsilon_1^{*} \varUpsilon_2^{*}} [(M^\ast x +m^\ast_q) (M^\ast (1-x) +m^\ast_{\bar{q}})-\bfk^2] \, , \nonumber \\
\Phi_K(x,\bfk,\downarrow,\downarrow) &=& \frac{1}{\sqrt{2}\varUpsilon_1^{*} \varUpsilon_2^{*}} [(M^\ast x +m^\ast_q) k_2^r - (M^\ast (1-x) +m^\ast_{\bar{q}}) k_1^r] \, . 
\label{SpinWfns}
\ee 
These spin wave functions must satisfy the normalization condition
\be 
\sum_{\lambda_1 \lambda_2} \Phi^\ast_K (x,\bfk,\lambda_1,\lambda_2) \, \Phi_K (x,\bfk,\lambda_1,\lambda_2) = 1 \,.
\ee 
The expressions of coefficients, written in Eq. (\ref{SpinWfns}) are expressed as $\varUpsilon_1^{*} = \sqrt{(M^\ast x +m_q^{\ast2})^2 + \bfk^2}$ and $\varUpsilon_2^{*} = \sqrt{(M^\ast (1-x) +m_{\bar{q}}^{\ast2})^2 + \bfk^2}$. The terms $k_{1(2)}^{r,l}$ represent $k_{1(2)}^{r,l}=k_{1(2)}^1\pm k_{1(2)}^2$ and the quantity $M^\ast$ satisfies the condition
\be 
M^{\ast2} = \frac{m_q^{\ast2} + \bfk^2}{x} + \frac{m_{\bar{q}}^{\ast2} + \bfk^2}{1-x} \, .
\ee  
We have used the Brodsky-Huang-Lepage prescription \cite{Yu:2007hp, Xiao:2002iv} as momentum space wave function that can be written as 
\be 
\varphi_K (x,\bfk)=\mathcal{A} \, exp \,\Biggl[-\frac{ \frac{m_{q}^{\ast2} + \bfk^2}{x} + \frac{m_{\bar{q}}^{\ast2} + \bfk^2}{1-x}}{8 \beta^2} - \frac{(m_q^{\ast2} - m_{\bar{q}}^{\ast2})^2}{8 \beta^2 \, \bigg( \frac{m_q^{\ast2} + \bfk^2}{x} + \frac{m_{\bar{q}}^{\ast2} + \bfk^2}{1-x}\bigg)}\Biggr] \, . \label{momspace}
\ee 
This wave function choice automatically provides a cutoff for non-zero transversal momentum $|\bfk|$ as it falls off exponentially in the endpoints $x\rightarrow0$ and $x\rightarrow1$. The quantity $\mathcal{A}$ in the above equation is defined as $\mathcal{A}= A \, exp \, \big[\frac{m_q^{\ast2} + m_{\bar{q}}^{\ast2}}{8 \beta^2}\big]$ with $A$ and $\beta$ as the normalization constant and harmonic scale parameter, respectively. The momentum space wave function is normalized as

\begin{eqnarray}
	\int \frac{{d x}~ d^2 \bfk}{2 (2 \pi)^3} \, |\varphi_K (x,\bfk)|^2 =1 \, .
\end{eqnarray}
The fraction of longitudinal momentum carried by the $j$th parton in the free space, i.e., $x_j$ is expressed as
\begin{align}
	x_j = \frac{k_j^+}{P^+} = 
	\frac{k_j^0 + k_j^3}{P^0+P^3},
	\label{Eq_xfrac_free}
\end{align}
where $k_j^0 = E_j$ and $P^0 = E_\pi$ correspond to the energies of $j$th quark and kaon respectively.
For the study of in-medium properties of kaon,
Eq. (\ref{Eq_xfrac_free}) has been modified to 
\cite{Arifi:2023jfe}
\begin{align}
	x_j^* = \frac{k_j^{*+}}{P^{*+}} = 
	\frac{k_j^{*0} + k_j^{*3}}{P^{*0}+P^{*3}}.
	\label{Eq_xfrac_med1}
\end{align}
In the CQMF framework, the vector fields $\omega$ and $\rho$ contribute to the  determination of in-medium values of $k_j^{*0}$ and $P^{*0}$. In detail, a discussion about the in-medium longitudinal momentum fraction $x^\ast$ has been presented in our previous work \cite{Puhan:2024xdq} and Refs. \cite{Arifi:2023jfe, Hutauruk:2019ipp}. However, to streamline our calculations, we have focused on $x$ only.
\subsection{\label{distribution}Distribution amplitude}
The information about the DAs can be obtained from the exclusive processes at large momentum transfer. On integrating LCWFs over the transverse momentum $\bfk$, we can get the light-cone distributions. For pseudo-scalar mesons, the correlation of DAs can be defined as \cite{Li:2017mlw, Bodwin:2006dm,Choi:2007yu}
\be 
\langle 0|\bar{\vartheta}(z) \gamma^+ \gamma_5 \vartheta(-z)|K (P^+,\bfP) \rangle = i k^+ \kappa \int_{0}^{1} dx \, e^{i(x-1/2) k^+ z^- } \phi (x) \bigg|_{z^+,\bfz=0} \, ,
\ee
where $\vartheta$ denotes the quark field operator with $\kappa$ as the decay constant. On substituting the kaon Fock state from Eq. (\ref{eqnq}) and quark field operators, the medium-modified DAs $\phi^\ast (x,m^\ast_u, m^\ast_{\bar {d}})$ (henceforth to be referred as $\phi(x)$) can be expressed in terms of the LCWFs as 
\be 
\frac{\kappa^\ast}{2 \sqrt{2 N_c}} \phi (x)=\frac{1}{\sqrt{2x(1-x)}} \int \frac{d^2 \bfk}{16 \pi^3} [\psi_K (x,\bfk,\uparrow,\downarrow)-\psi_K (x,\bfk,\downarrow,\uparrow)] \, ,\label{DA}
\ee 
where $N_c=3$ denotes the number of colours of a quark flavour, and $\kappa^*$ represents the in-medium decay constant. The kaon DA is normalized as 
\be 
\int_{0}^{1} dx \, \phi(x)=1 \, .
\ee
\subsection{\label{parton}Parton distribution functions}
PDFs provide the probability of locating the valence quark with longitudinal momentum fraction $x$ in a kaon. The correlator of PDF at a fixed light-front time can be expressed as \cite{Maji:2016yqo}
\be 
f_q(x)=\frac{1}{2} \int \frac{dz^-}{4 \pi} e^{i k^+ z_- /2} \langle K(P^+,\bfP;S)|\bar{\vartheta}(0) \Gamma \vartheta (z^-)|K(P^+,\bfP;S) \rangle |_{z^+, \bfz=0} \, ,
\ee
where $\Gamma=\gamma^+$ and spin for the pseudoscalar meson $K$ is $S=0$. On substituting the meson state in the above equation from Eq. (\ref{eqnq}), the unpolarized PDF $f_q(x,m_q^\ast,m_{\bar{q}}^\ast)$ (henceforth to be referred as $f_q(x)$) can be expressed in terms of overlap form of LCWFs as 
\be 
f_q(x)&=&\int \frac{d^2 \bfk}{16 \pi^3} \big[|\psi_K (x,\bfk,\uparrow,\uparrow)|^2 +|\psi_K (x,\bfk,\uparrow,\downarrow)|^2  + |\psi_K (x,\bfk,\downarrow,\uparrow)|^2 \nonumber \\ &+& |\psi_K (x,\bfk,\downarrow,\downarrow)|^2 \big] \, .
\ee
The explicit form of valence quark PDF is written as 
\be
f_q(x)&=& \int \frac{d^2 \bfk}{16 \pi^3} \bigg[\big((x {M}^*+m^*_{q})((1-x){M}^*+m^*_{\bar q})-\textbf{k}^2_\perp\big)^2 
+ \big({M}^*+ m^*_q+m^*_{\bar q}\big)^2\bigg] \nonumber \\ &\times& \frac{\mid \varphi_K^*(x,\textbf{k}_\perp)\mid^2}{\varUpsilon^{*2}_1 \varUpsilon^{*2}_2}. \label{pdf}
\ee
PDFs corresponding to an antiquark of kaon can be obtained by using the $f_q(1-x)$ function. Both vacuum and in-medium unpolarized PDF in Eq. (\ref{pdf}) obey the PDF sum rule \cite{Kaur:2020vkq,Puhan:2023ekt,Puhan:2023hio}
\be
\int d x f_q(x) =\int d x  f_{\bar{q}}(1-x) =1,
\nonumber \\
\int d x [xf_q(x) + (1-x) f_q(x)] = 1.
\ee
Non-perturbative aspects of the QCD can be studied through Mellin moments of quark PDFs. Higher-order Mellin moments correspond to the quark densities with respect to different momentum fractions, whereas the first moment defines the fraction of an average momentum carried by the quark from its parent meson. The Mellin moments of unpolarized kaon PDF can be expressed as \cite{Lu:2023yna}
\be
\langle x^n\rangle=\frac{\int dx ~ x^n f_q(x)}{\int dx ~ f_q(x)}.
\ee

\section{\label{results}Results and Discussions}

In this section, we will discuss the medium modifications of DAs and PDFs of kaon calculated using the LCQM model with medium-modified masses of quarks in asymmetric nuclear matter obtained using Eq. (\ref{qmass}) within the CQMF model. Table \ref{tab:1} lists different parameters used to solve the equations of motion of the CQMF model to obtain the density and temperature dependence of the scalar fields $\sigma,\zeta,$ and $\delta$ through Eqs. (\ref{Eq_sigma_eq1})-(\ref{Eq_rho_field}). The model parameters are adjusted to match the vacuum values of various mesons masses and nuclear matter saturation properties \cite{Wang:2002pza}. In addition, we have also studied the effect of isospin asymmetry on the kaon and antikaon doublets $K$ and $\bar{K}$, respectively. Finally, we have calculated the medium-modified Mellin moments using the medium-modified PDFs of valence $u$ and $s$ quarks of kaon. 

\begin{table}[h]
\footnotesize{
\centering
\begin{tabular}{|c|c|c|c|c|c|c|c|c|c|}
\hline
$k_0$           & $k_1$          & $k_2$          & $k_3$         & $k_4$         & $g_s$         & $\rm{g_v}$          & $\rm{g_4}$           & $\xi$          & $\rho_0$(fm$^{-3}$)                            \\ \hline
4.94                 & 2.12                & -10.16              & -5.38              & -0.06              & 3.85               &  9.14                & 37.4                 & 6/33               & 0.16                                  \\ \hline
$\sigma_0$ (MeV) & $\zeta_0$(MeV)  & $\chi_0$(MeV)   & $m_\pi$(MeV)  & $f_\pi$(MeV)  & $m_K$(MeV)    & $f_K$(MeV)     & $m_\omega$(MeV) & $m_\phi$(MeV)  & $m_\rho$( MeV)                   \\ \hline
-93                  & -96.87             & 254.6               & 139                & 93                 & 496                & 115                 & 783                  & 1020                & 783                                   \\ \hline
$g_{\sigma}^u$ = $g_{\sigma}^d$ & $g_{\sigma}^s$ = $g_{\zeta}^u$ = $g_{\zeta}^d$ & $g_{\zeta}^s$ & $g_{\delta}^u$ = $g_{\delta}^d$  & $g_{\delta}^s$ & $g^u_{\omega}$ = $g^d_{\omega}$ & $g^s_{\omega}$ & $g^u_{\rho}$   = $g^d_{\rho}$ &   $g^s_{\rho}$  & $g^p_{\sigma}$=$g^n_{\sigma}$  \\ \hline
2.72                & 0                      & 3.847               & 2.72                & 0                   &      3.23          &    0 &  2.72    & 0      &          6.64           \\ \hline
 $g^p_{\zeta}$=$g^n_{\zeta}$ &  $g^p_{\delta}$=$g^n_{\delta}$   &$g^p_{\omega}$=$g^n_{\omega}$   &$g^p_{\rho}$=$g^n_{\rho}$     & & & & & &                           \\ \hline
0 &   2.72    &    9.69     &   8.886  & & & & & &         \\ \hline
\end{tabular}
\caption{The list of parameters used in the present work \cite{Wang:2001jw}.}
\label{tab:1}
}
\end{table}

\begin{figure*}
\centering
\begin{minipage}[c]{0.98\textwidth}
(a)\includegraphics[width=4.7cm]{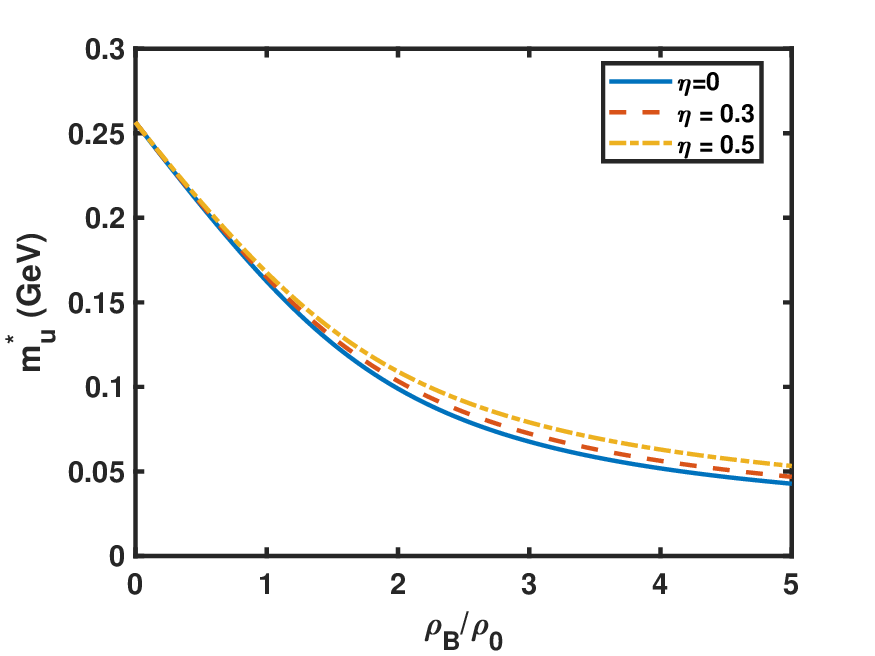}
\hspace{0.02cm}
(b)\includegraphics[width=4.7cm]{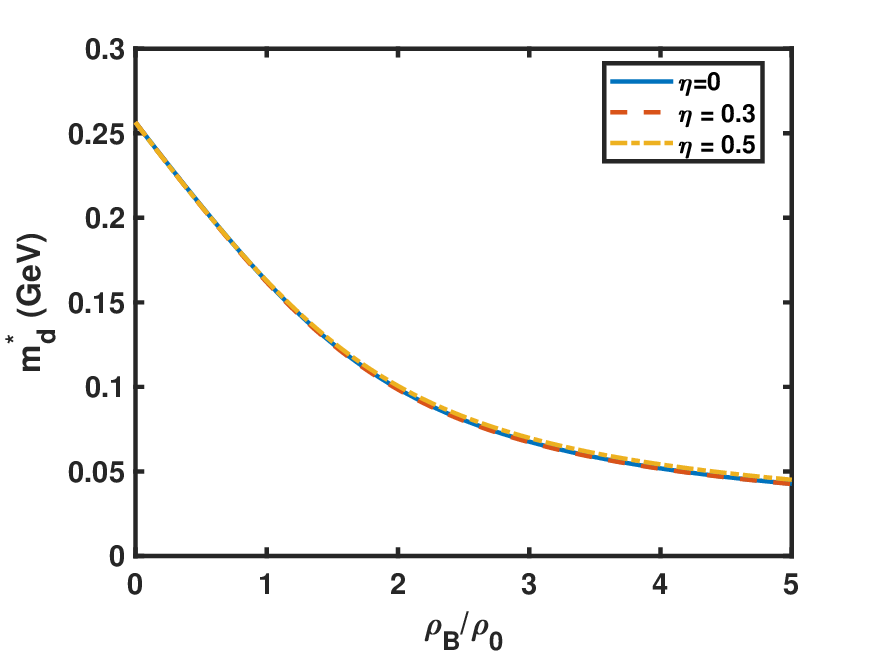}
\hspace{0.02cm}	
(c)\includegraphics[width=4.7cm]{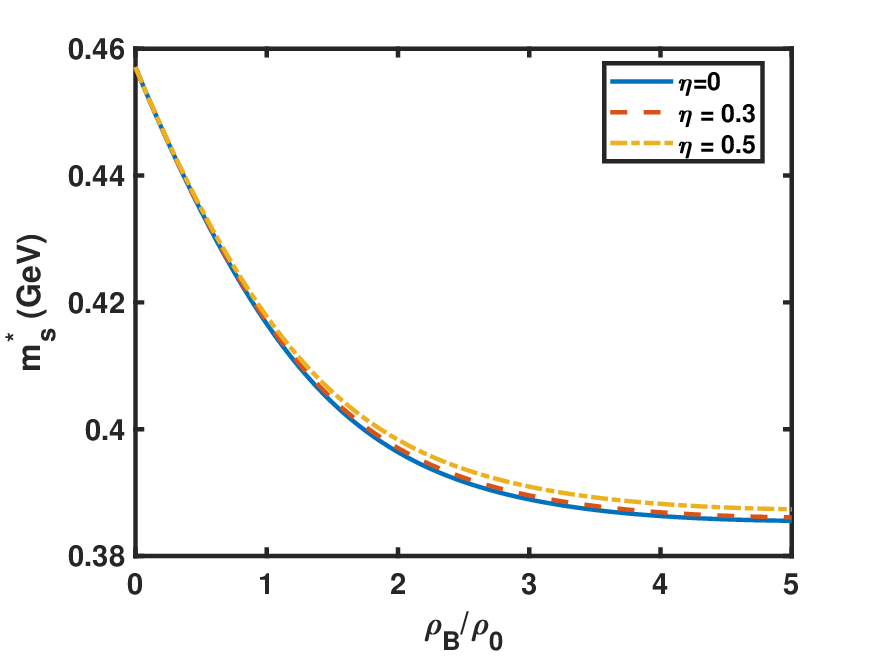}
\hspace{0.02cm}	
\end{minipage}
\caption{\label{fig1mu} (Color online) Effective mass of $u,d$, and $s$ quarks as a function of baryonic density for different values of asymmetry $\eta = 0,0.3$, and 0.5.}
\end{figure*}

\begin{figure*}
\includegraphics[width=8cm]{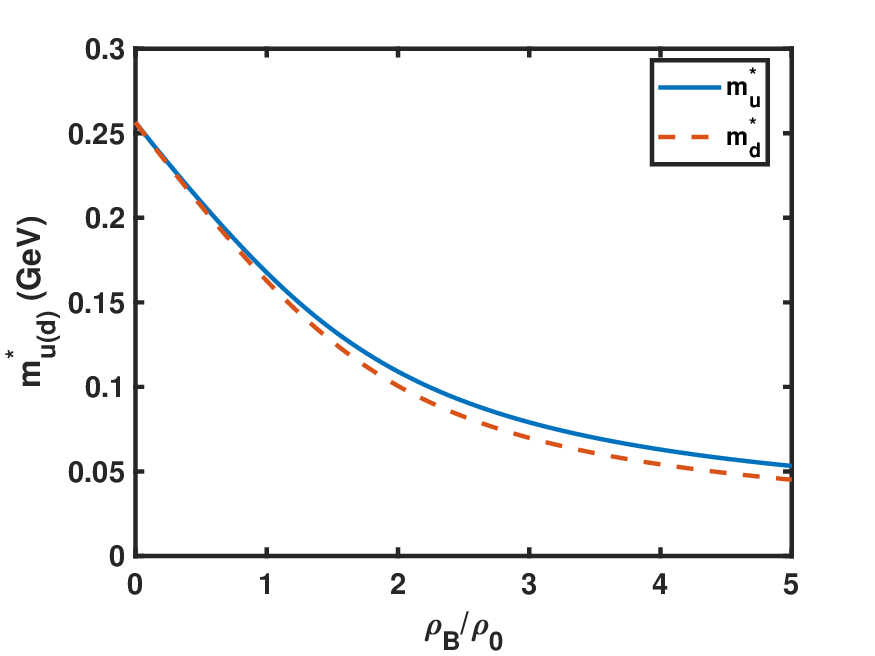}
\caption{\label{fig2mud} (Color online) Effective mass of $u$ and $d$ quarks as a function of baryonic density for zero temperature and $\eta=0.5$.}
\end{figure*}

\begin{figure*}
\centering
\begin{minipage}[c]{0.98\textwidth}
(a)\includegraphics[width=4.7cm]{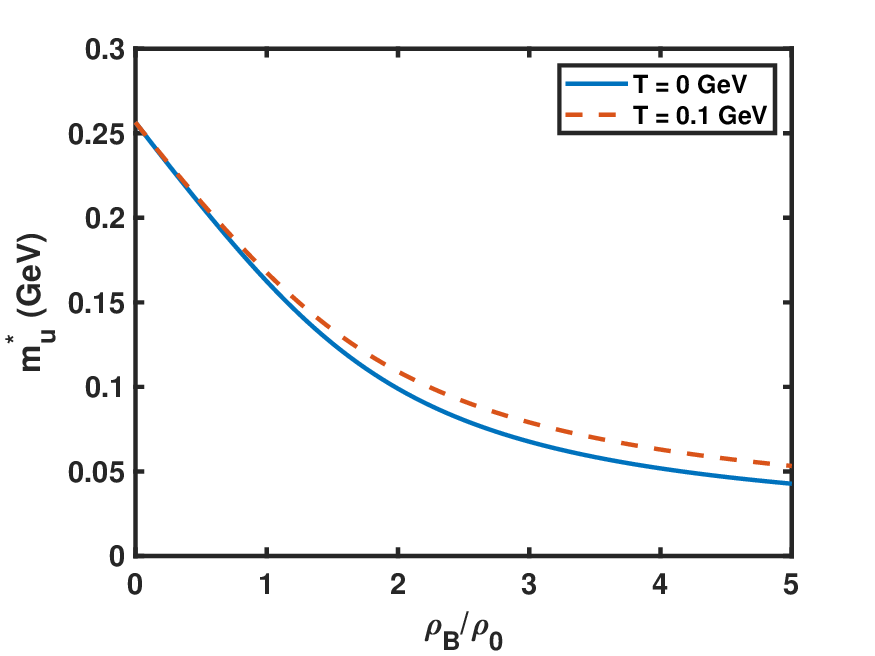}
\hspace{0.02cm}
(b)\includegraphics[width=4.7cm]{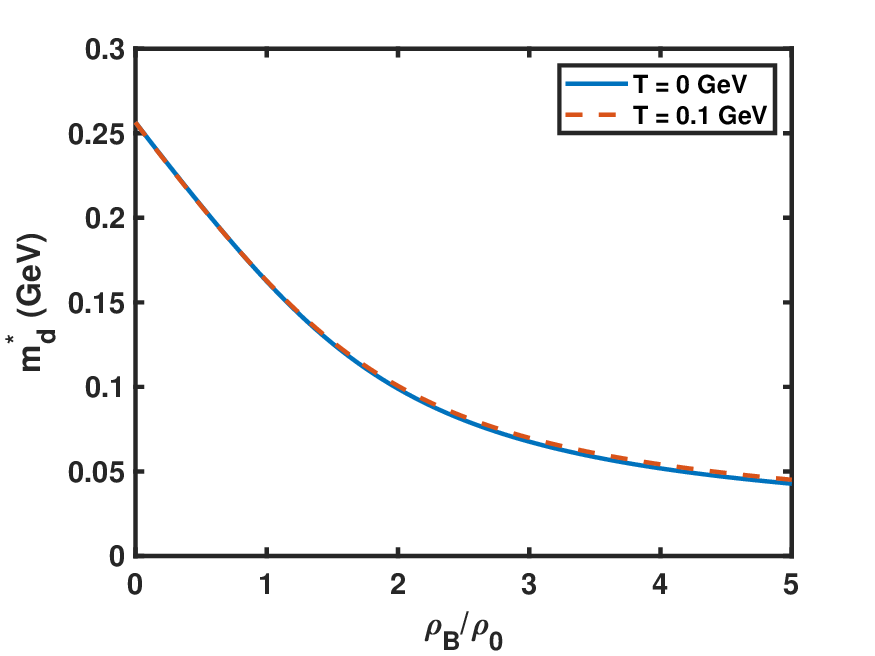}
\hspace{0.02cm}	
(c)\includegraphics[width=4.7cm]{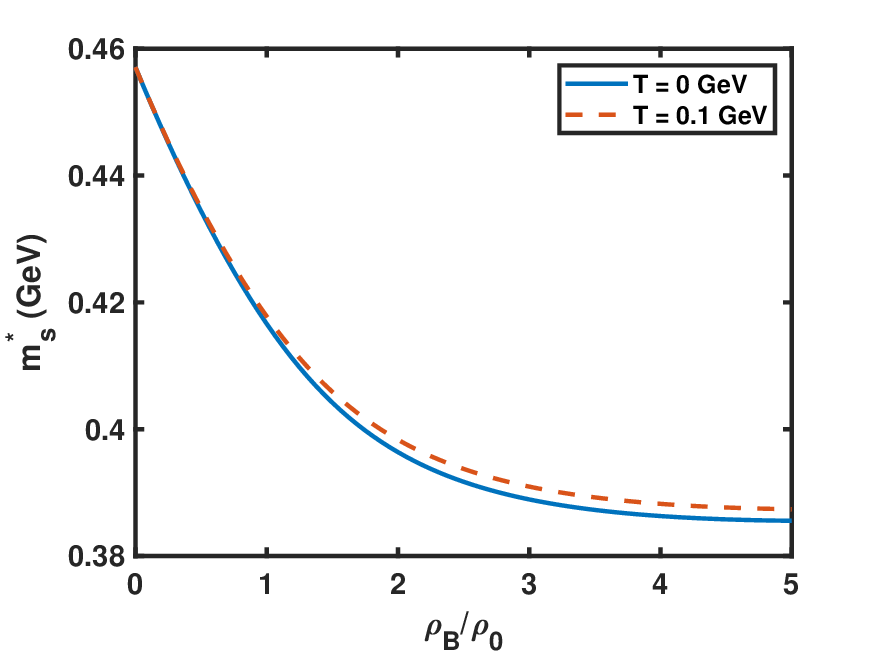}
\hspace{0.02cm}	
\end{minipage}
\caption{\label{fig3mut} (Color online) Effective mass of $u,d$, and $s$ quarks as a function of baryonic density for different values of temperature $T=0$ and 0.1 GeV and $\eta=0$.}
\end{figure*}

To study the effective masses of quarks in the nuclear medium with finite isospin asymmetry, in Fig. \ref{fig1mu}, we have plotted the effective masses of $u,d,$ and $s$ quarks as a function of baryonic density ratio ($\rho_B/\rho_0$) for different values of isospin asymmetry $\eta=0,0.3$, and 0.5. Effective masses of quark decrease with increasing baryonic density for all flavors. At finite value of $\eta$, the effective masses of $u$ and $s$ quarks are considerably affected at higher values of $\rho_B/\rho_0$ which is clear from Fig. \ref{fig1mu} (a) and \ref{fig1mu}(c). However, the effective mass of $d$ quarks remains largely unaffected for finite asymmetry in the medium. As the isospin asymmetry in the medium increases ($\eta=0.5$), the distinction between the $u$ and $d$ flavors increases which is clear from Fig. \ref{fig2mud}. At $T=0$ and for a fixed value of $\rho_B/\rho_0$, the effective mass of the $d$ quark is higher than that of the $u$ quark. Additionally, we observe that higher medium density increases the influence of isospin asymmetry. The density variations of the effective quark masses of all three flavours are shown for symmetric nuclear matter at different values of temperature $T=0$ and 0.1 GeV in Fig. \ref{fig3mut}. It is observed that the rise in temperature of the medium leads to an increase in the effective quark masses for all three flavors. To calculate in-medium DAs and PDFs of kaon in LCQM, we have used the effective quark masses $m^*_u,m^*_d,$ and $m^*_s$ and a fixed harmonic scale $\beta=0.405$ as input parameters.

\begin{figure*}
\centering
\begin{minipage}[c]{0.98\textwidth}
(a)\includegraphics[width=7.4cm]{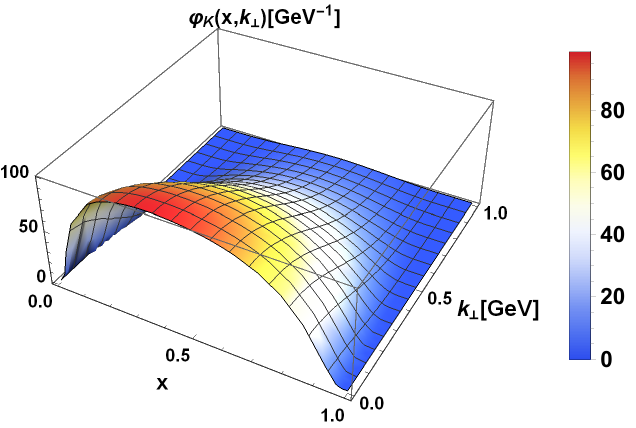}
\hspace{0.03cm}
(b)\includegraphics[width=7.4cm]{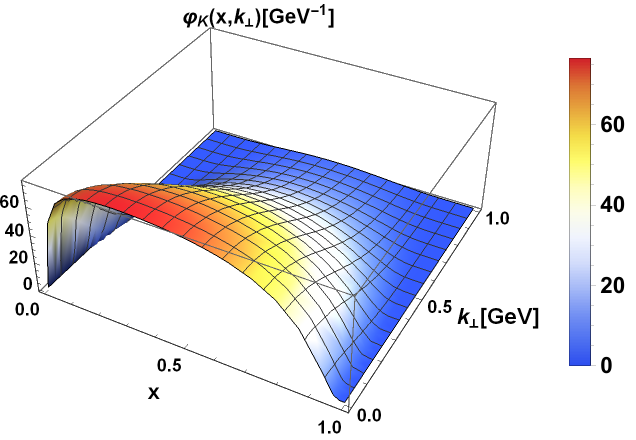}
\hspace{0.03cm}	
(c)\includegraphics[width=7.4cm]{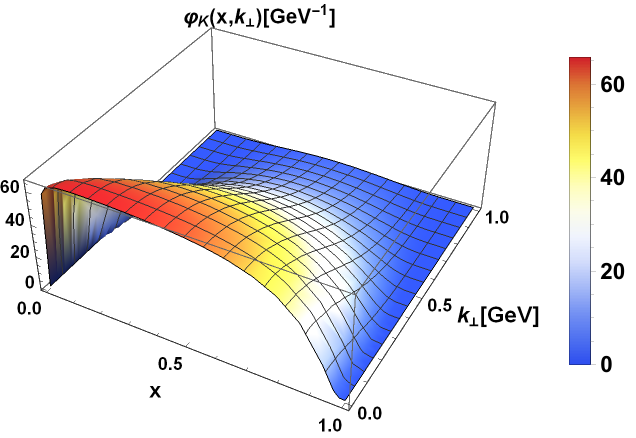}
\hspace{0.03cm}
(d)\includegraphics[width=7.4cm]{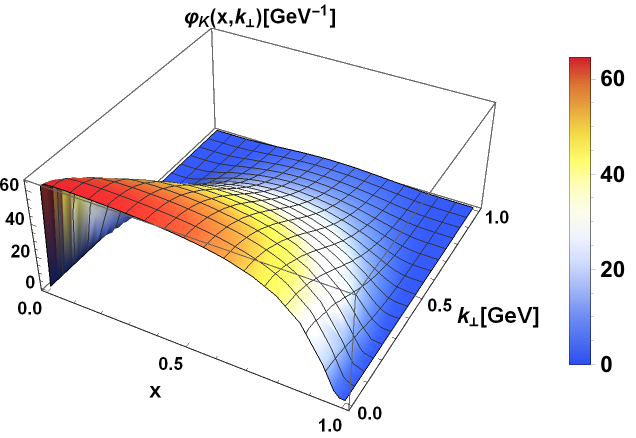}
\hspace{0.03cm}	
\end{minipage}
\caption{\label{fig4WF3D} (Color online) Dependence of light-cone momentum space wave function $\varphi_K (x,\bfk)$ on longitudinal momentum fraction $x$ and transverse momentum $\bfk$  for baryonic density ratio (a) $\rho_B/\rho_0=0$, (b) $\rho_B/\rho_0=1$, (c) $\rho_B/\rho_0=3$ and (d) $\rho_B/\rho_0=5$ in symmetric nuclear medium at zero temperature.}
\end{figure*}

\begin{figure}
\centering
\begin{minipage}[c]{0.98\textwidth}
\includegraphics[width=7.5cm]{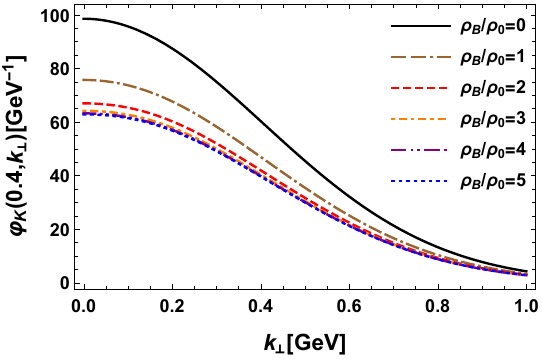}
\hspace{0.03cm}		
\end{minipage}
\caption{\label{fig5Wavefn} (Color online) Comparison of vacuum and in-medium light-cone momentum space wave functions with respect to $\bfk$ for different values of baryonic densities at fixed longitudinal momentum fraction $x=0.4$ for the case of $T=0$ and $\eta=0$.}
\end{figure}

In Fig. \ref{fig4WF3D}, we have plotted the three-dimensional kaon momentum space wave function $\varphi_K (x,\bfk)$ as a function of transverse momentum $\bfk$ and longitudinal momentum fraction $x$ at baryonic densities $\rho_B/\rho_0=0,1,3,$ and 5 (in units of $\rho_0$). For symmetric nuclear matter ($\eta = 0$) and zero temperature, we observe that the momentum space wave function of kaon is symmetric around $x=0.5$. However, it becomes asymmetric and shifts slightly towards lower $x$ with increasing baryon density. In order to study the influence of baryon density on momentum space wave function, we have presented in Fig. \ref{fig5Wavefn} the two-dimensional plot as a function of transverse momentum $\bfk$ at a given longitudinal momentum fraction $x = 0.4$ for specific values of baryon density. The momentum space wave function exhibits a significant reduction at lower transverse momentum values as the value of baryonic density is increased.

\begin{figure*}
\centering
\begin{minipage}[c]{0.98\textwidth}
\includegraphics[width=7.5cm]{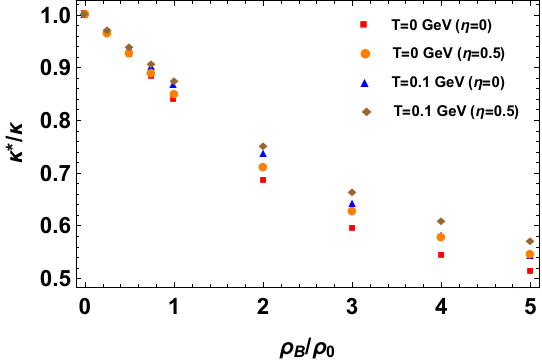}
\hspace{0.03cm}	
\end{minipage}
\caption{\label{fig6DecayConstant} (Color online) Decay constant ratios vs baryon density for different values of temperature $T$ and asymmetry $\eta$.}
\end{figure*}

To analyse the effect of the nuclear medium on the decay constant of kaon, in Fig. \ref{fig6DecayConstant} we have plotted the ratio of decay constant in medium to free space as a function of baryonic density ratio $\rho_B/\rho_0$ for different values of temperature $T=0,0.1$ GeV and asymmetry parameter $\eta = 0,0.5$. We found that $\kappa^*/\kappa$ becomes less than 1 as the baryonic density increases, i.e., the medium effects appear only at higher values of baryonic density, and they disappear at lower densities as $\kappa^*/\kappa\rightarrow1$. However, the values of $\kappa^*/\kappa$ are found to be significant in our model calculations and in Ref. \cite{Yabusaki:2023zin} than computed in Ref. \cite{Hutauruk:2019ipp}. For instance, the value of $\kappa^*/\kappa$ at $\rho_B=\rho_0$ is found to be $0.838$ in our case and $0.967$ in Ref. \cite{Hutauruk:2019ipp}. The values of the decay constant ratio at different baryonic densities are compared with Refs. \cite{Yabusaki:2023zin,Hutauruk:2019ipp} in Table \ref{table_kk}. The in-medium effects of temperature on the decay constant ratio are more pronounced than those of asymmetry at lower densities. However, as $\rho_B/\rho_0$ is increased to higher values, we observe that the asymmetry effects become more dominating than the temperature effects. As $\rho_B/\rho_0\rightarrow0$, the in-medium effects on the decay constant vanishes and become independent of both $T$ and $\eta$.
\begin{table}
    \centering
   \begin{tabular}{|c|c|c|c|}
    \hline
     & \multicolumn{3}{c|}{Ratio of decay constant ($\kappa^*/\kappa$)}  \\
     \cline{2-4}
    $~~~~\rho_B/\rho_0~~~~$ & ~~This work~~ & ~~~~Ref.\cite{Hutauruk:2019ipp}~~~~ & ~~~~Ref.\cite{Yabusaki:2023zin}~~~~ \\
    \hline
    0 & 1 & 1 & 1  \\
    \hline
    0.25 & 0.963 & 0.952 & 1  \\
    \hline
    0.50 & 0.924 & 0.906 & 0.989  \\
    \hline
    0.75 & 0.881 & 0.863 & 0.978  \\
    \hline
    1 & 0.838 & 0.822 & 0.967  \\
    \hline
    2 & 0.684 & 0.717 & -  \\
    \hline
    3 & 0.592 & 0.647 & -  \\
    \hline
\end{tabular}

    \caption{Comparison of computed decay constant ratios at different baryonic densities with Ref. \cite{Hutauruk:2019ipp, Yabusaki:2023zin} for zero temperature and symmetric medium.}
    \label{table_kk}
\end{table}

\begin{figure*}
\centering
\begin{minipage}[c]{0.98\textwidth}
(a)\includegraphics[width=7.4cm]{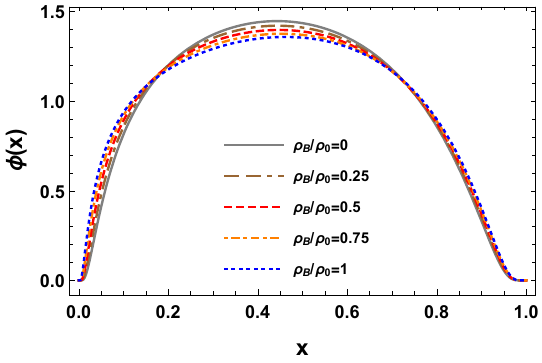}
\hspace{0.03cm}
(b)\includegraphics[width=7.4cm]{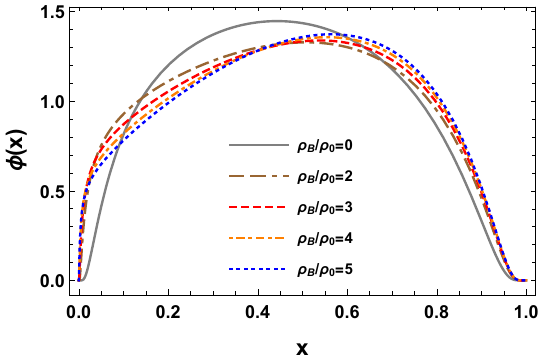}
\hspace{0.03cm}		
\end{minipage}
\caption{\label{fig7DAdensity} (Color online) Comparison of distribution amplitudes of kaon in vacuum and medium as a function of longitudinal momentum fraction in symmetric nuclear matter at zero temperature. The left panel represents the comparison of vacuum distribution with baryon density up to $1$, and the right panel represents the comparison of vacuum distribution with baryon density above $1$ as well.}
\end{figure*}

The in-medium DAs of kaon $\phi(x)$ in the symmetric nuclear matter ($\eta=0$) are plotted as a function of longitudinal momentum fraction $x$ for a wide range of relative baryonic density $\rho_B/\rho_0 = 0$ to 1 in Fig. \ref{fig7DAdensity} (a) and $\rho_B/\rho_0 = 0$ to 5 in Fig. \ref{fig7DAdensity} (b). We observe that $\phi(x)$ becomes slightly asymmetric and increases as $\rho_B\rightarrow\rho_0$ for lower $x$. However, for $x$ in the range 0.1 to 0.6, a contrasting decrease of $\phi(x)$ is observed with increasing baryonic density. This may be a consequence of the reduction in the effective quark masses, indicating partial restoration of chiral symmetry. 
A similar kind of behaviour is also found in Ref. \cite{Arifi:2023jfe}. The in-medium DAs of kaon become highly asymmetric as the value of $\rho_B/\rho_0$ becomes higher than 1, as is evidenced in Fig. \ref{fig7DAdensity} (b). In the near endpoint regions $x\rightarrow0$ and $x\rightarrow1$, the value of $\phi(x)$ is enhanced for higher densities as compared to its vacuum value at $\rho_B/\rho_0=0$. However, 
we observe that $\phi(x)$ is suppressed for higher densities at values of $0.1<x<0.6$. 
Also, the peak of kaon in-medium DAs is shifted towards a higher longitudinal momentum fraction for a denser medium. This may be due to the higher effective mass difference between light ($u$) and strange ($\bar{s}$) quark in the medium. Additionally, we found that the values of $\phi(x)$ decreases (increases) with increasing baryonic density from $\rho_B/\rho_0 = 2$ to 5 at longitudinal momentum fraction $0.1<x<0.4$ ($0.4<x<0.8$). While the effect of increasing baryonic density ($\rho_B/\rho_0 = 2\rightarrow5$) disappears in both near-endpoint regions. 

\begin{figure*}
\centering
\begin{minipage}[c]{0.98\textwidth}
(a)\includegraphics[width=7.4cm]{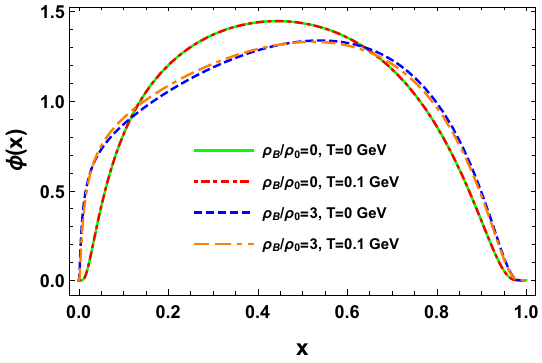}
\hspace{0.03cm}
(b)\includegraphics[width=7.4cm]{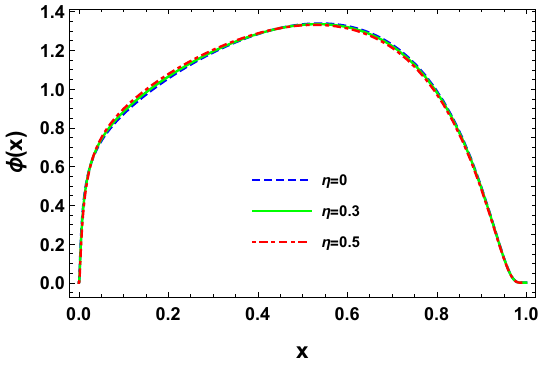}
\hspace{0.03cm}		
\end{minipage}
\caption{\label{fig8DAtempEta} (Color online) Comparison of distribution amplitudes of kaon as a function of longitudinal momentum fraction (a) for different values of temperature $T$ at baryonic density $\rho_B/\rho_0=0$ and $3$ and (b) for different values of isospin asymmetry $\eta$ for fixed baryonic density $\rho_B/\rho_0=3$.}
\end{figure*}

To study the influence of temperature on the kaon DAs, we have plotted $\phi(x)$ as a function of longitudinal momentum fraction $x$ at baryonic density $\rho_B/\rho_0=0$ and $3$ for temperatures $T=0$ and 0.1 GeV in Fig. \ref{fig8DAtempEta} (a). At zero density, the increase in temperature of the medium seems to have a negligible effect on the kaon DAs. However, at $\rho_B/\rho_0=3$, $\phi(x)$ increases with the rising temperature of the medium in the range $0.1<x<0.5$. Beyond these values, the in-medium DAs of kaon are suppressed with increasing $T$. The temperature of the medium has no observable effect on the medium-modified DAs of kaon in the near endpoints of $x$. The effect of the isospin asymmetry in the medium is studied in Fig. \ref{fig8DAtempEta} (b). Similar to the temperature effect, the isospin asymmetry of the medium enhances the DAs of kaon for around $0.1<x<0.4$ and suppresses beyond that.

\begin{figure*}
\centering
\begin{minipage}[c]{0.98\textwidth}
(a)\includegraphics[width=7.4cm]{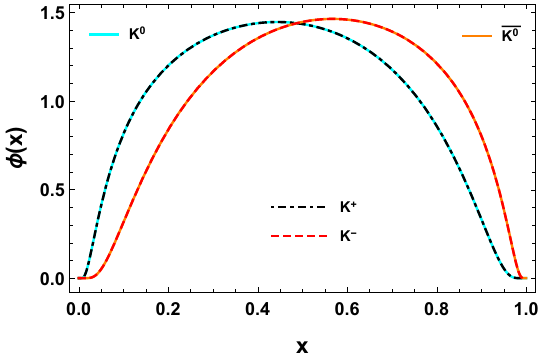}
\hspace{0.03cm}
(b)\includegraphics[width=7.4cm]{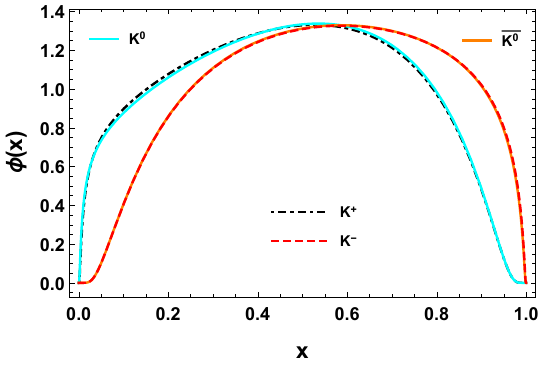}
\hspace{0.03cm}		
\end{minipage}
\caption{\label{fig9DA_Doublets} (Color online) Comparison of distribution amplitudes of kaon doublets as a function of longitudinal momentum fraction at temperature $T=0$ GeV for (a) baryonic density $\rho_B/\rho_0=0$ in symmetric medium $\eta=0$ and (b) baryonic density $\rho_B/\rho_0=3$ in asymmetric medium with $\eta=0.5$.}
\end{figure*}

To study the difference in the in-medium DAs of the kaon doublet $K \left(\begin{array}{c} K^{+} \\ K^{0} \end{array} \right),$ and antikaon doublet $\bar{K}$ ($K^-, \bar{K}^0$) we have plotted them as a function of longitudinal momentum fraction $x$ at zero temperature for symmetric nuclear medium ($\eta=0$) and zero baryonic density in subplot Fig. \ref{fig9DA_Doublets} (a) and asymmetric medium ($\eta=0.5$) at baryonic density $\rho_B = 3\rho_0$ in subplot Fig. \ref{fig9DA_Doublets} (b). We observe that the DAs of the kaons ($K^+$ and ${K^0}$) are strikingly different from those of antikaons ($K^-$ and $\bar{K^0}$), even in the vacuum (Fig. \ref{fig9DA_Doublets} (a)). The dissimilarity in the DAs of kaons and antikaons is due to the difference in their quark content. For example, $K^+$ contains a $u$ quark and an $\bar{s}$ antiquark, while its antikaon $K^-$ contains a $\bar{u}$ antiquark and an $s$ quark. According to Eq. (\ref{qmass}), the in-medium mass of $\bar{u}$ is equivalent to the in-medium mass of $d$ quark as $I^{\bar{u}}_{3} = I^{d}_{3} = -\frac{1}{2}$. The in-medium masses are used in Eqs. (\ref{SpinWfns}) and (\ref{momspace}) to calculate the LCWFs, which are then used to compute the in-medium DAs according to Eq. (\ref{DA}). In the case of $K^+$ the mass of quark $m^*_q = m^*_u$ and that of antiquark is $m^*_{\bar{q}} = m^*_{\bar{s}}=m^*_s$. Whereas, for $K^-$, the mass of quark is $m^*_q = m^*_s$ and the antiquark mass is $m^*_{\bar{q}}=m^*_{\bar{u}}=m^*_d$. This difference in the quark contents in kaons and antikaons causes modification in the LCWFs which in turn is responsible for the differences observed in the DAs. We can also see that the DAs of the members of the isospin doublets ($K$ and $\bar{K}$) are exactly the same in vacuum (Fig. \ref{fig9DA_Doublets} (a)). This is because in the vacuum $m^*_u=m^*_d$. When we increase the asymmetry in the medium to $\eta=0.5$ and density to $3\rho_0$, we find that the behaviour of the DAs of $K$ is similar as in Fig. \ref{fig7DAdensity} (b). However, the DAs of $\bar{K}$ show an opposite shift. The introduction of the asymmetry in the medium seems to have a very vague effect on the DAs of both kaon and antikaon isospin doublets, similar to Fig. \ref{fig8DAtempEta} (b).

\begin{figure*}
\centering
\begin{minipage}[c]{0.98\textwidth}
(a)\includegraphics[width=7.4cm]{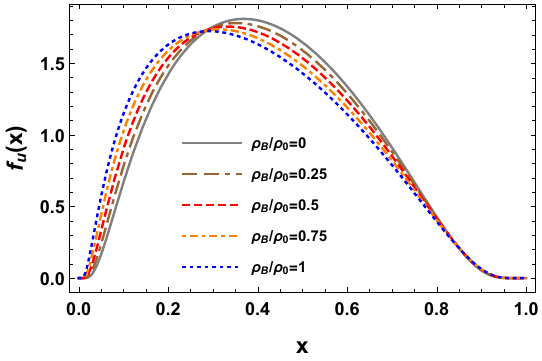}
\hspace{0.03cm}
(b)\includegraphics[width=7.4cm]{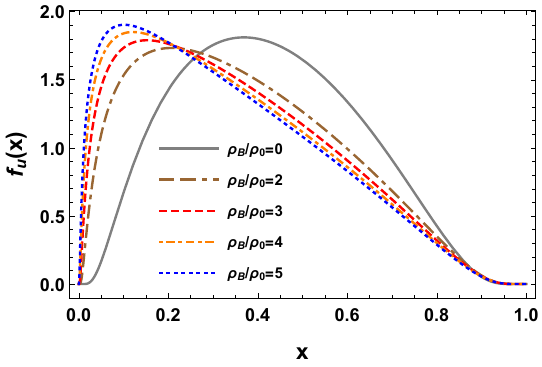}
\hspace{0.03cm}	
(c)\includegraphics[width=7.4cm]{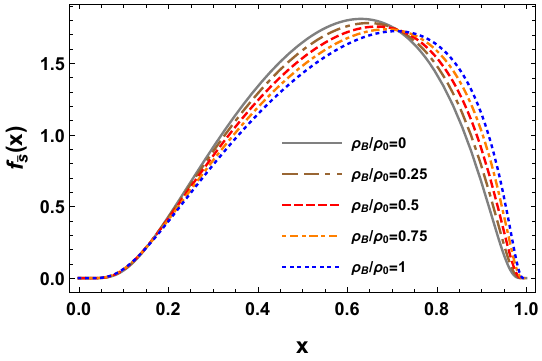}
\hspace{0.03cm}
(d)\includegraphics[width=7.4cm]{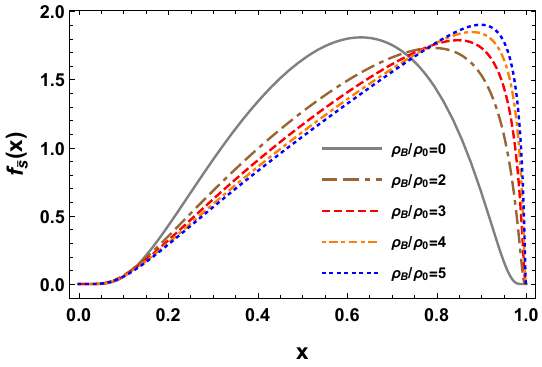}
\hspace{0.03cm}	
\end{minipage}
\caption{\label{fig10PDFs} (Color online) Comparison of vacuum and in-medium parton distribution functions for valence $u$ quark (Row $1$) and $\bar{s}$ antiquark (Row $2$) of kaon as a function of longitudinal momentum fraction in symmetric nuclear matter at zero temperature. The left panel represents the comparison of vacuum distribution with baryon density up to $1$, and right panel represents the comparison of vacuum distribution with baryon density above $1$.}
\end{figure*}

The in-medium PDFs of the valence $u$ quark of kaon can be calculated using the medium-modified masses of quarks obtained from Eq. (\ref{qmass}) and the LCWFs mentioned in Eqs. (\ref{SpinWfns}) and (\ref{momspace}) in Eq. (\ref{pdf})  while that of the valence $\bar{s}$ antiquark can be obtained through $f_q(1-x)$ as mentioned in Sec. \ref{parton}. The valence PDF of the $u$ quark of kaon in a symmetric nuclear medium as a function of longitudinal momentum fraction $x$ for a range of densities up to $\rho_0$ is shown in Fig. \ref{fig10PDFs} (a). We find that increasing the baryonic density shifts the position of the peak towards lower longitudinal momentum fraction, indicating that $u$ quark carries lesser momentum in a denser medium. Additionally, we note that the amplitude of the PDF is enhanced at lower $x$ and suppressed at higher $x$ as $\rho_B$ is increased. In Fig. \ref{fig10PDFs} (b), the variation of the PDF of $u$ quark is shown for values up to $\rho_B/\rho_0 = 5$. We see that the peak gets shifted to even lower values of $x$, again implying that the valence $u$ quark of kaon carries less longitudinal momentum fraction in a dense medium. This means that the valence $\bar{s}$ antiquark of kaon must carry a higher longitudinal momentum fraction at higher values of $\rho_B$, and this is evident in Figs. \ref{fig10PDFs} (c) and \ref{fig10PDFs} (d), where we have plotted the PDFs of the $\bar{s}$ antiquark of kaon as a function of $x$ for a range of densities.

\begin{figure*}
\centering
\begin{minipage}[c]{0.98\textwidth}
(a)\includegraphics[width=7.4cm]{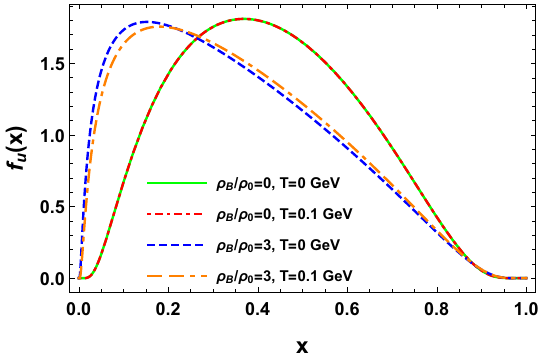}
\hspace{0.03cm}
(b)\includegraphics[width=7.4cm]{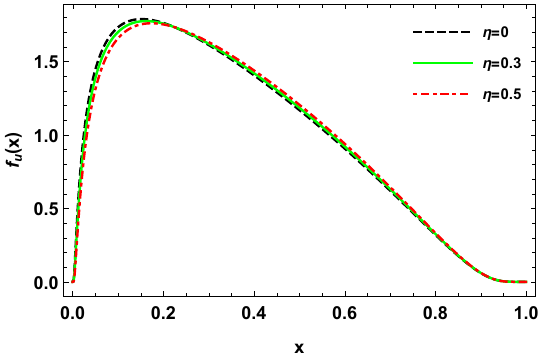}
\hspace{0.03cm}	
(c)\includegraphics[width=7.4cm]{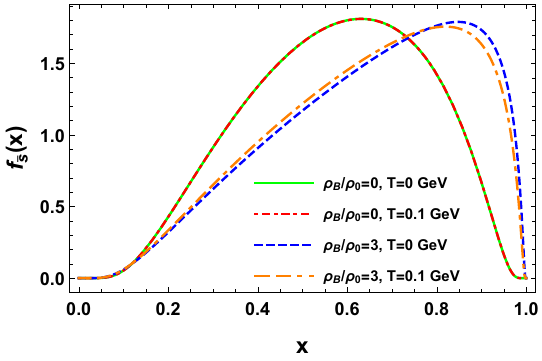}
\hspace{0.03cm}
(d)\includegraphics[width=7.4cm]{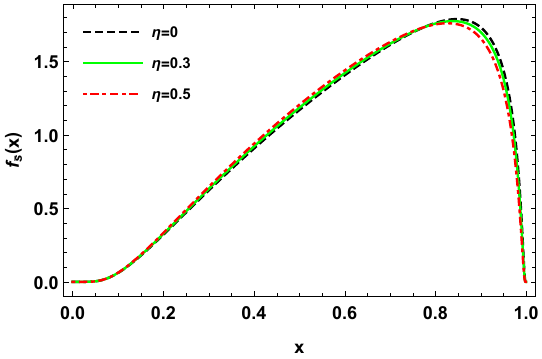}
\hspace{0.03cm}	
\end{minipage}
\caption{\label{fig11PDFtemp} (Color online) Comparison of parton distribution functions for valence $u$ quark (Row $1$) and $\bar{s}$ antiquark (Row $2$) of kaon as a function of longitudinal momentum fraction. The left panel represents the temperature dependence for vacuum and in-medium distribution at $\eta=0$, and the right panel represents the asymmetry dependence of in-medium distribution for baryonic density $\rho_B/\rho_0=3$ at $T=0$.}
\end{figure*}

The temperature dependence of the PDFs of valence quarks (antiquarks) of kaon is studied through Fig. \ref{fig11PDFtemp}, where in Fig. \ref{fig11PDFtemp} (a), we have demonstrated the valence $u$ quark PDF of kaon as a function of $x$ at baryon densities $\rho_B =0 $ and $3\rho_0$ and compared the results at temperature $T=0$ and 0.1 GeV. At $\rho_B=3\rho_0$, we find that in contrast to the impact of the density, increasing the temperature from 0 to 0.1 GeV shifts the peak position of the PDF towards higher $x$. Also, the amplitude of the PDF decreases with an increase in temperature for longitudinal momentum fraction $x < 0.2$. However, in the range $0.2<x<0.8$, the amplitude increases with rising temperature. We note that the impact of temperature of the medium on the PDF vanishes when $\rho_B/\rho_0=0$. The valence $\bar{s}$ antiquark PDF of kaon is shown as a function of longitudinal momentum fraction $x$ for temperatures $T=0$ and 0.1 GeV and densities $\rho_B=0$ and $3\rho_0$ in Fig. \ref{fig11PDFtemp} (c). The impact of the temperature of the medium on the behaviour of $\bar{s}$ antiquark PDF is found to be opposite to that of the $u$ quark, i.e., the peak position of the PDF is moved towards lower longitudinal momentum fraction $x$ with an increase in temperature from 0 to 0.1 GeV. The amplitude of the $\bar{s}$ antiquark PDF is found to increase (decrease) with rising medium temperature for longitudinal momentum fraction around $0.2<x<0.8$ ($x>0.8$). Similar to the $u$ quark PDF, we see that the influence of temperature on the $\bar{s}$ antiquark PDF disappears when $\rho_B/\rho_0$ becomes zero. To study the impact of the isospin asymmetry of the medium, we plotted the PDFs of $u$ and $\bar{s}$ as a function of longitudinal momentum fraction $x$ for isospin asymmetry $\eta =0,0.3,$ and 0.5 at temperature $T=0$ in Figs. \ref{fig11PDFtemp} (b) and \ref{fig11PDFtemp} (d), respectively. Increasing the asymmetry in the medium seems to have a similar effect as the rising temperature on the PDF of $u$ ($\bar{s}$), i.e., the peak position is shifted towards a higher (lower) $x$ value. Hence, we conclude that if a kaon is immersed in an asymmetric dense nuclear medium, then its constituent valence $u$ quark ($\bar{s}$ antiquark) will carry slightly more (less) fraction of longitudinal momentum.

\begin{figure*}
\centering
\begin{minipage}[c]{0.98\textwidth}
(a)\includegraphics[width=7.4cm]{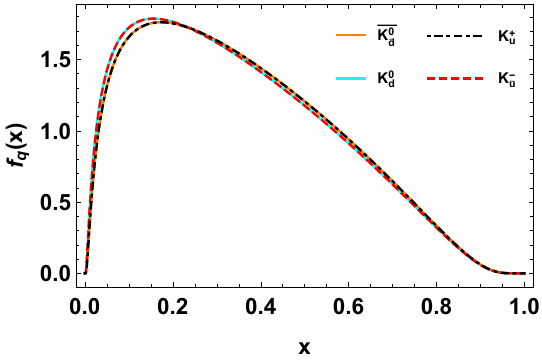}
\hspace{0.03cm}
(b)\includegraphics[width=7.4cm]{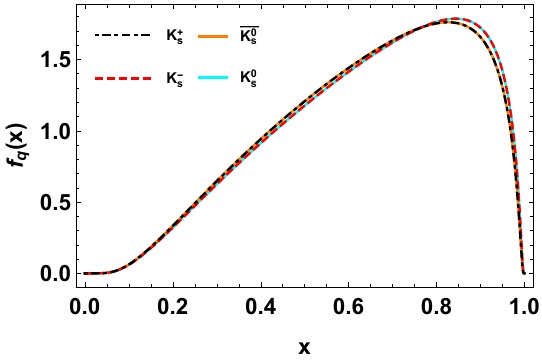}
\hspace{0.03cm}	
\hspace{0.03cm}	
\end{minipage}
\caption{\label{fig12PDFtemp} (Color online) Comparison of parton distribution functions of valence quarks of kaon and antikaon doublets as a function of longitudinal momentum fraction for baryonic density $\rho_B/\rho_0=3$ at isospin asymmetry $\eta=0.5$ and temperature $T=0$ GeV.}
\end{figure*}

In Fig. \ref{fig12PDFtemp}, we have shown the longitudinal momentum fraction dependence of the medium-modified PDFs of valence quarks and antiquarks of kaon and antikaon doublets at zero temperature and baryonic density and isospin asymmetry fixed at $\rho_B=3\rho_0$ and $\eta = 0.5$, respectively. Similar to the DAs of the isospin doublets in Fig. \ref{fig9DA_Doublets}, we observe that the PDFs of the valence quarks/antiquarks of isospin doublets are exactly the same. However, the PDFs of the valence quark/antiquark of the kaon doublet are marginally different from those of the antikaon doublet. Again, this small variation is because the PDFs depend upon the LCWFs according to Eq. (\ref{pdf}), which are different for kaons and antikaons, as mentioned in the discussion of Fig. \ref{fig9DA_Doublets}.

\begin{figure*}
\centering
\begin{minipage}[c]{0.98\textwidth}
(a)\includegraphics[width=7.4cm]{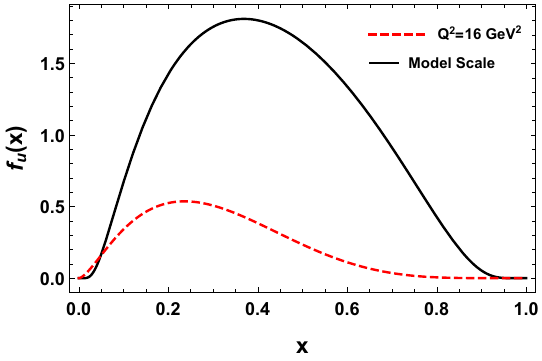}
\hspace{0.03cm}
(b)\includegraphics[width=7.4cm]{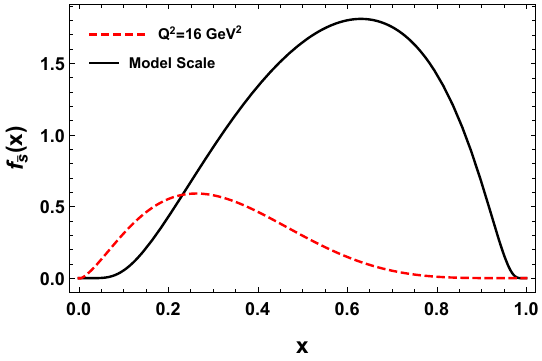}
\hspace{0.03cm}		
\end{minipage}
\caption{\label{fig13Mod-Evo} (Color online) Vacuum parton distribution function for valence (a) $u$ quark and (b) $\bar{s}$ antiquark of kaon at model scale ($Q^2=0.235$ GeV$^2$) compared with evolved distribution at $Q^2=16$ GeV$^2$.}
\end{figure*}

\begin{figure*}
\centering
\begin{minipage}[c]{0.98\textwidth}
(a)\includegraphics[width=7.4cm]{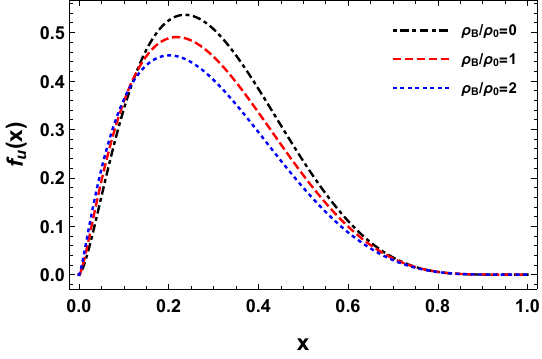}
\hspace{0.03cm}
(b)\includegraphics[width=7.4cm]{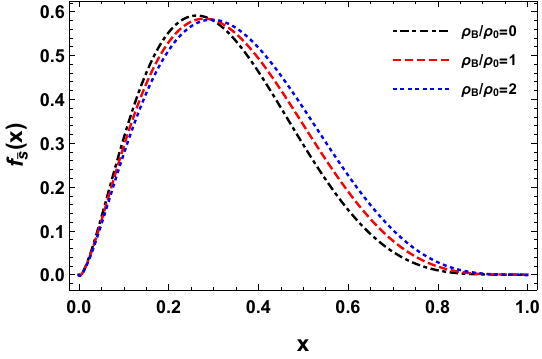}
\hspace{0.03cm}		
\end{minipage}
\caption{\label{fig14Evo} (Color online) Evolved parton distribution functions for valence (a) $u$ quark and (b) $\bar{s}$ antiquark of kaon in symmetric nuclear medium at zero temperature for fixed values of baryonic density.}
\end{figure*}

To examine the behaviour of the valence quark PDFs in the high $Q^2$ region, we have evolved them using next-to-leading order (NLO) Dokshitzer-Gribov-Lipatov-Altarelli-Parisi (DGLAP)
evolution equations from our model scale $Q^2=0.235$ GeV$^2$ to $Q^2=16$ GeV$^2$ \cite{Altarelli:1977zs}. The evolved PDFs are shown in Fig. \ref{fig13Mod-Evo}. We find that the amplitude of the evolved PDFs is distinctly small, and the peak position is shifted towards a lower longitudinal momentum fraction of $x$ compared to the model-scale PDFs. This may be because of the reduction of the effective masses of quarks at high energies. The baryonic density dependence of the PDFs is studied in Fig. \ref{fig14Evo}, where we have shown the variation of the evolved PDFs as a function of longitudinal momentum fraction $x$ for baryonic densities $\rho_B/\rho_0=0,1$, and 2 at zero temperature and asymmetry. We find that the peak position of the valence $u$ PDF is shifted towards lower longitudinal momentum fraction as $\rho_B$ is increased while the valence $\bar{s}$ PDF is moved towards higher $x$ values. This is similar to the results obtained for our model scale (Fig. \ref{fig10PDFs}). The trend of shifting of peaks with the increase in baryon density over the domain of $x$ for the case of both $u$ as well as $s$ quark is consistent with the results obtained in Ref. \cite{Hutauruk:2019ipp}, but the amplitudes of distributions fall down in our calculations. The evolved in-medium PDFs in our model are compared to those in the NJL model at $\rho_B=\rho_0$ and zero temperature. The distributions of the valence quarks are quite different in both models. However, the peak position of the PDFs is almost the same.

We compare the vacuum and in-medium Mellin moments of valence $u$ quark and $\bar{s}$ antiquark kaon PDF at the model scale up to $n = 10$ for symmetric matter and zero temperature. The vacuum and in-medium Mellin moments of both $u$ and $\bar{s}$ are the same when $n=0$. However, as $n$ becomes greater than zero, the Mellin moment of the $u$ quark increases while that of the $\bar{s}$ antiquark decreases. We note that as $n$ is further increased to higher values, the density dependence in the Mellin moments of the $u$ quark disappears.

\section{\label{summ}Summary}
In the present work, we computed the mass of dynamical quarks in an isospin asymmetric nuclear medium at zero as well as finite temperature within the framework of the chiral SU(3) quark mean field (CQMF) model. We have then examined the in-medium properties of kaons in asymmetric nuclear matter in the Light-cone quark model (LCQM) utilizing the in-medium quark properties acquired in the CQMF model as inputs. In particular, we have calculated the weak decay constant, distribution amplitudes (DAs), and valence quark distribution functions (PDFs) of kaon. The CQMF model modifies the constituent quark masses in isospin asymmetric matter using scalar fields $\sigma$, $\zeta,$ and $\delta$. The isospin asymmetry is introduced in the model through the dimensionless $\eta$ parameter. The dynamic spin effect has been incorporated into the LCQM, which is used to calculate kaon properties. We have derived the DAs and PDFs in the overlap representation of light-cone wave functions (LCWFs) within LCQM by analyzing the quark-quark correlation function for spin-0 mesons, utilizing medium-modified effective quark masses as input parameters.

The DAs of kaon have been calculated both in vacuum as well as at finite value of temperature $T$, baryonic density $\rho_B$, and isospin asymmetry $\eta$. We found that with increasing baryon density, the DA of kaon is enhanced in the near endpoint $x$ region while it is suppressed in the range $0.1<x<0.6$. Also, increasing the temperature and isospin asymmetry of the medium results in enhancement of the DA value of kaon for $0.1<x<0.4$ and suppression beyond that. However, the effects of density are observed to be more significant than those of temperature and isospin asymmetry. We have also investigated the in-medium DAs of the kaon doublet $K= \left(\begin{array}{c} K^{+} \\ K^{0} \end{array} \right),$ and antikaon doublet $\bar{K}$= ($K^-, \bar{K}^0$). The DAs of kaons $K$ are found to be quite different from those of antikaons $\bar{K}$ due to differences in their quark contents. However, medium-modified DAs of the isospin doublets $K$ and $\bar{K}$ are found to be identical. 

The valence $u$ quark and $\bar{s}$ antiquark PDFs of kaon have also been computed. The constituent $u$ quark is found to carry lesser fraction of longitudinal momentum in a denser nuclear medium, and consequently, the $\bar{s}$ antiquark carries more fraction of longitudinal momentum as compared to that in the vacuum. In contrast, the valence $u$ quark ($\bar{s}$ antiquark) is observed to carry more (less) fraction of the longitudinal momentum of kaon for a hot, dense, and asymmetric medium than the vacuum value. The PDFs of the kaon doublet are found to be slightly different than those of the antikaon doublet in the isospin asymmetric nuclear medium. In order to study the valence quark PDFs in the high $Q^2$ region, the valence PDFs of kaon were evolved to $Q^2 = 16$ GeV$^2$. The amplitude of the evolved PDFs is quite smaller than the model-scale PDFs. Also, the peak position of the $u$ quark PDF is shifted towards lower $x$ while that of $\bar{s}$ antiquark is shifted towards higher $x$. Finally, we have computed the Mellin moments of valence quark of kaon up to $n=10$ using the PDFs. We observe that increasing the value of $n$ results in Mellin moment of valence $u$ quark to increase and that of the valence $\bar{s}$ antiquark to decrease. These modifications in the properties of kaon may be due to the reduction of quark masses implying partial restoration of chiral symmetry.

\section*{ACKNOWLEDGMENTS}

A.K. sincerely acknowledge Anusandhan-National Research Foundation (ANRF), Government of India for funding of the
research project under the Science and Engineering Research Board-Core Research Grant (SERB-CRG) scheme (File No. CRG/2023/000557).
H.D. would like to thank  the Science and Engineering Research Board, Anusandhan-National Research Foundation (ANRF), Government of India under the SERB-POWER Fellowship scheme (Ref No. SPF/2023/000116) for financial support.

\section{Reference}

\bibliography{Mine}

\begin{figure*}
\centering
\begin{minipage}[c]{0.98\textwidth}
(a)\includegraphics[width=7.4cm]{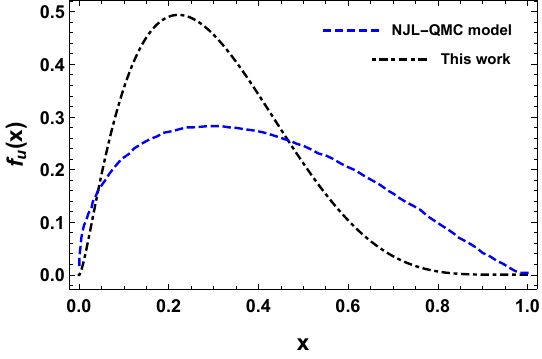}
\hspace{0.03cm}
(b)\includegraphics[width=7.4cm]{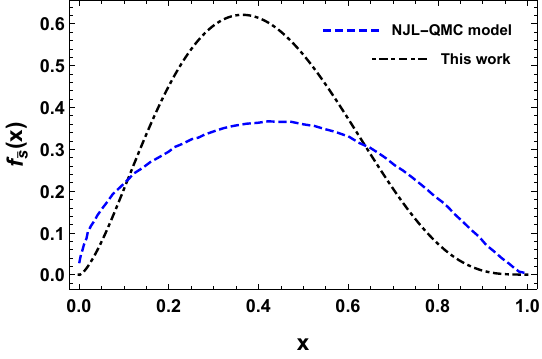}
\hspace{0.03cm}		
\end{minipage}
\caption{\label{fig15CompEVO} (Color online) Evolved parton distribution functions for valence (a) $u$ quark and (b) $\bar{s}$ antiquark of kaon in symmetric nuclear medium at baryonic density $\rho_B/\rho_0=1$ for temperature $T=0$ GeV compared with NJL-QMC model results \cite{Hutauruk:2019ipp}.}
\end{figure*} 
\begin{figure*}
\centering
\begin{minipage}[c]{0.98\textwidth}
\includegraphics[width=7.5cm]{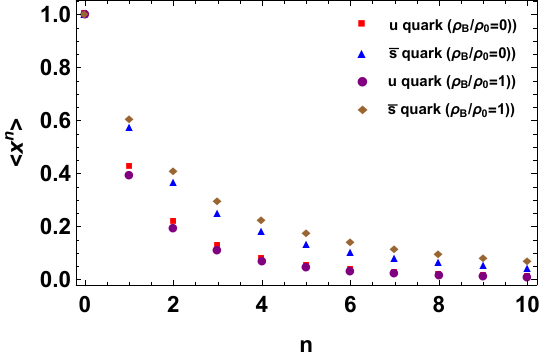}
\hspace{0.03cm}
\end{minipage}
\caption{\label{fig16} (Color online) Comparison of vacuum and in-medium Mellin moments of valence quark of kaon for $\rho_B/\rho_0=0$ and $\rho_B=\rho_0$ in symmetric nuclear medium at zero temperature.}
\end{figure*} 

 \end{document}